\documentclass[fleqn,usenatbib]{mnras}
\usepackage{dcolumn}
\usepackage{graphicx}
\usepackage{url}
\usepackage{color}
\usepackage[T1]{fontenc}
\usepackage{ae,aecompl}
\usepackage{pdflscape}
\usepackage[normalem]{ulem}

\usepackage{verbatim}

\newcommand{\cm}{cm$^{-1}$}

\newcommand{\ai}{\textit{ab initio}}

\newcommand{\etal}{\textit{et al.}}
\newcommand{\eqref}[1]{(\ref{#1})}
\newcommand{\Dh}{${\mathcal D}_{3{\rm h}}$}

\newcommand{\TROVE}{{\sc TROVE}}

\newcommand{\sothree}{SO$_3$}
\newcommand{\p}{^\prime}
\newcommand{\pp}{^{\prime\prime}}

\title[ExoMol XVII: Line list for SO$_3$]{ExoMol molecular line lists  - XVII The
rotation-vibration spectrum of hot SO$_3$}

\author[D.S. Underwood et al.]{Daniel S. Underwood$^{1}$,  Sergei N. Yurchenko$^{1}$, Jonathan
Tennyson$^{1}$\thanks{Email: j.tennyson@ucl.ac.uk},
\newauthor Ahmed F. Al-Refaie$^{1}$,
S{\o}nnik Clausen$^2$ and Alexander Fateev$^{2}$
\\
$^{1}$Department of Physics and Astronomy, University College London, London
WC1E 6BT, UK\\
$^{2}$Technical University of Denmark, Department of Chemical and Biochemical
Engineering, Frederiksborgvej 399,
4000 Roskilde, Denmark}
\date{Accepted XXXX. Received XXXX; in original form XXXX}
\pubyear{2016}

\begin{document}
\label{firstpage}
\pagerange{\pageref{firstpage}--\pageref{lastpage}}

\maketitle

\begin{abstract}
Sulphur trioxide (SO$_3$) is a trace species in the atmospheres of the Earth and Venus,
as well as well as being an industrial product and an environmental pollutant.
A variational line list for $^{32}$S$^{16}$O$_{3}$, named UYT2, is presented containing 21 billion vibration-rotation
transitions. UYT2 can be used to model
infrared spectra of SO$_3$ at wavelengths longwards of 2 $\mu$m ($\nu < 5000$ cm$^{-1}$)
for temperatures up to 800 K. Infrared absorption cross sections are also
recorded at 300 and 500 C are used to validate the
UYT2 line list.  The intensities in UYT2 are scaled to match the measured cross sections.
The line list is made available in
electronic form as supplementary data to this article and at
\url{www.exomol.com}.

\end{abstract}


\section{Introduction}

SO$_3$ known to exist naturally in the Earth's atmosphere; its main
source being volcanic emissions and hot springs
\citep{05MiKrGrAn.SO3} but it also plays 
role in the formation
of acid rain. The oxidisation of SO$_2$ to SO$_3$ in the atmosphere,
followed by subsequent rapid reaction with water vapour results in the
production of sulphuric acid (H$_2$SO$_4$) \citep{ 85CaLaKo.SO3}
with many adverse
environmental effects \citep{11VaGoHa.SO3,94KoJaWo.SO3,04SrMiEr.SO3}.
So SO$_3$ is a natural product whose concentration in the atmosphere
is significantly enhanced by human activity, particularly as a
byproduct of industrialisation.  SO$_3$ is observed in
the products of combustion processes
\citep{04SrMiEr.SO3,14HiMexx.SO3} and selective
catalytic reduction (SCR) units, where the presence of both is undesirable
within flue gas chambers in large quantities, as well as other industrial
exhausts \citep{05RaHeSoOa.SO3,12FlVaAnBr.SO3}. The control of these outputs is
therefore of great importance.
The spectroscopic study of  sulphur
oxides can also provide insight into the history of the Earth's
atmosphere \citep{13WhXiHu.SO2}. All this means that
observation of SO$_3$ spectra and hence
concentrations provide a useful tool for understanding 
geological processes and controlling polution.

Sulphur oxide chemistry has been observed in a variety of
astrophysical settings. Within the solar system, SO$_3$ is a
constituent of the atmosphere of Venus
\citep{83CrReRa.SO3,10ZhLiMoBe.SO3,13ZhLiMa.SO3}. Although SO$_3$ has
yet to be observed outside our solar system, it needs to be considered
alongside other sulphur oxides namely, sulphur monoxide (SO) and SO$_2$,
which are well-known in several astronomical environments
\citep{90NaEsSk.SO2,92PeBexx.SO,03MaMaMa.SO2,05MaMaMa.SO2,12BeMoBe.SO2,06ViLoFe.SO2,13BeMuMe.SO2,13AdEdZi.SO2,15KhViMu.SO2}.
SO$_3$ chemistry has been considered in a number of enviroments
including giant planets, brown dwarfs, and dwarf stars
\citep{06ViLoFe.SO2}.  Unlike SO and SO$_2$, SO$_3$ is a symmetric
species with no permanent dipole moment making it hard to detect in
the interstellar medium.  In practice, the identification of SO$_3$ in
the infrared is hindered by the presence of interfering SO$_2$ where
both species are found simultaneously; a number of their spectral
features overlap, particularly the $\nu_3$ bands of both molecules in
the 1300 - 1400 \cm\ (7.4 $\mu$m) region. From this point of view
SO$_2$ can also be seen as a spectral `weed' with respect to the
detection of SO$_3$. An understanding of the spectroscopic behaviour
of both of these molecules within the same spectral window is
therefore required to be able to correctly identify each species
independently. In this context we note that a number of line lists are
available for SO$_2$ isotopologues
\citep{14HuScLe.SO2,16HuScLe.SO2,jt635}; of particular relevence is
the recent hot ExoAmes line list of \citet{jt635}.

The experimental spectroscopic studies of SO$_3$ have significant gaps, notably
the absence of any measurement of absolute line intensities in the infrared.
This may be attributed to its vigorous chemical reactivity which make
measurements difficult.  SO$_3$
is a symmetric planar molecule with
equilibrium S-O bond lengths of 1.41732 \AA\ and inter-bond angles of
120$^{\circ}$ \citep{89OrEsMa.SO3}, described by \Dh(M) symmetry. The $\nu_1$, $\nu_2$,
$\nu_3$ and $\nu_4$ fundamental frequencies are attributed to the totally
symmetric stretch at 1064.9 \cm \citep{02BaChMa.SO3}, the symmetric bend at 497.5
\cm, and the asymmetric stretching and bending modes at 1391.5 and 530.1 \cm,
respectively \citep{03ShBlSa.SO3}.

The infrared and coherent anti-stokes vibration-rotation spectra of a number of
isotopologues of SO$_3$ have been extensively investigated in a series of papers
by Maki and co-workers
\citep{73KaMaDo.SO3,89OrEsMa.SO3,01ChVuMa.SO3,01MaBlSa.SO3,02BaChMa.SO3,
03ShBlSa.SO3,04MaBlSa.SO3}, reassessing and confirming fundamental constants and
frequencies. 18 bands were analysed based on an empirical fitting to
effective Hamiltonian models, yielding rovibrational constants and energy levels
assigned by appropriate vibrational and rotational quantum numbers.
Some temperature-dependent infrared cross sections are also available from laboratory studies
\citep{13GrFaNi.SO2,14GrFaCl.SO3}, and we present new measurements in this work.
Unlike all other measurements of SO$_3$ spectra, these cross sections are absolute. However,
assigned spectra represented by line lists allow for the modelling of both absorption
and emission spectra in different environments.

The ``forbidden'' rotational spectrum, for which centrifugal distortions can
induce transitions, was investigated for the first time by \citet{91MeSuDr.SO3} using
microwave Fourier-transform spectroscopy. Assignments for 25
transitions were made, as well as the determination of a number of rotational
constants, including the only direct measurement of the $C_0$ rotational
constant. The work was analysed and extended theoretically \citep{jt580}.

There have been a few studies on the ultraviolet spectrum of SO$_3$ by
\citet{36FaGoxx.SO3}, and \citet{81LeBrRi.SO3}, both
between 220 and 270 nm where overlap with SO$_2$ is small.
\citet{97BuMcxx.SO3} reported cross sections for the 195 to
330 nm range for the purposes of photolysis rate calculations of SO$_3$. All
measurements were taken at room-temperature, and neither reported assignments
for any of the bands, which exist as weak, diffuse vibrational band structures
superimposed on a continuous background. As such, the rovibronic behaviour of
SO$_3$ is much less well understood than for SO$_2$.

Prior to our studies, there was limited theoretical work on SO$_3$.
Early work on anharmonic force constants \citep{73DoHoMi.SO3,92FlRoTa}
for \sothree\ led to first accurate, fully \ai\ anharmonic quartic
force field computed by \citet{99Maxxxx.SO3}. There have been no
theoretical studies into the UV spectrum of SO$_3$.  As for the
experimental studies for SO$_3$, none of this work provided transition
intensities.  Our preliminary study for this project \citep{jt554},
which produced the \ai, room-temperature UYT line list, therefore
provides the first absolute transition intensities for \sothree. These
were used in the 2012 release of the HITRAN database \citep{jt557} to
scale the relative experimental measurements allowing \sothree\ to be
included in the database for the first time. As discussed below, the
present work suggest that these intensites may need to be
reconsidered.

The present study on SO$_3$ was performed as part of the ExoMol
project.  ExoMol aims to provide comprehensive line lists of molecular
transitions important for understanding hot atmospheres of exoplanets
and other bodies \citep{jt528}.  Besides the ExoAmes SO$_2$ line list
mentioned above, ExoMol has produced very extensive line lists for a
number of polyatomic species including methane (CH$_4$) \citep{jt564},
phosphine (PH$_3$) \citep{jt592}, formaldehyde (H$_2$CO) \citep{jt597},
hydrogen peroxide (HOOH) \citep{jt620}
and nitric acid (HNO$_3$) \citep{jt614}.  These line lists, all of
which contain about 10 billion distinct vibration-rotation
transitions, required the adoption of special computational procedures
to make their calculation tractable.  The UYT2 SO$_3$ line list
presented here is the largest computed so far with 21 billion lines;
as SO$_3$ is a system comprising four heavy atoms this meant considering
rotation states up to $J = 130$ as part of these calculations. These
calculations therefore required further enhancement of our
computational methods which are described below.  The lack of measured
SO$_3$ spectra at temperatures above 300~K on an absolute scale is
clearly a problem for validating our calculations. Here we present
infrared absorption cross sections for SO$_3$ measured at a range of
temperatures up to 500~C.

The next section describes our theoretical procedures; our experiments
are described in section 3.  Section 4 presents the UYT2 line list.
Section 5 compares the UYT2 line list with our measurements with a
particular emphasis on intensity comparisons. The final section gives
details on how to access the line list and our conclusions.

\section{Theoretical method}

To compute a variational line list requires three components
\citep{jt475}: a suitable potential energy surface (PES), dipole
moment surfaces (DMS), and a nuclear motion program. Variational
nuclear motion programs, which use basis functions to provide direct
solutions of the rotation-vibration Schr\"odinger equation for a given
PES, means that interactions between the levels associated with
different vibrational states and the associated intensity stealing
between these bands are automatically included in the calculation. In
particular, the use of exact kinetic energy operators means that how well
these effects are reproduced depends strongly on the PES used; the reader
is refered to a recent study by \citet{jt625} for a discussion of
this.  

Here the nuclear motion calculations are performed with the
flexible, polyatomic vibration-rotation nuclear motion program \TROVE\
\citep{07YuThJe.method}.  The \ai\ DMS surface was adopted unaltered
from our previous calculations \citep{jt554} (UYT); below we describe
refinement of the PES.  Both the \ai\ PES and DMS were computed at the
coupled-clusters level of theory (CCSD(T)-F12b) level of theory with
appropriate triple-$\zeta$ basis sets, aug-cc-pVTZ-F12 and
aug-cc-pV(T+d)Z-F12 for O and S, respectively.

The label F12 in the theoretical model denotes the use of explicitly
correlated functions which are designed to accelarate basis set convergence.
The F12b varient is an efficient F12 implementation due to \citet{07AdKnWe}.
Use of CCSD(T)-F12b methods have been shown to give improved vibrational
frequencies compared to standard CCSD(T) calculations \citep{14MaKexx}
but their use for intensity calculations remains relatively untested.
We return to this issue below.

\subsection{Refining the Potential Energy Surface}

The refinement of the \ai\  PES involved performing a least-squares fit
to empirical rovibrational energies or observed transition
frequencies.  The procedure follows that described elsewhere
\citep{jt503,11YaYuJe.H2CO,jt556,jt564} and is based on
adding a correction,
$\Delta V$, to the \ai\ UYT PES, which is represented by an expansion
\begin{eqnarray}
\Delta V
    &=&  \sum_{ijklmn}  \, \Delta f_{ijklmn} \xi_1^{i} \xi_2^{j} \xi_3^{k} \xi_4^{l} \xi_5^{m} \xi_6^{n} \, \
\label{e:PEF:morbid}
\end{eqnarray}
in terms of the same internal coordinates as UYT \citet{jt554}:
\begin{eqnarray}
\label{CurvStretc1}
   \xi_{k}  &=& 1-\exp( -a( r_k -r_{\rm e} ) ) \hbox{,} \;\; k=1,2,3, \\
   \label{CurvXi4a}
   \xi_{4}            &=& \frac{1}{\sqrt{6}} \left( 2 \alpha_{23} - \alpha_{13} - \alpha_{12} \right)\hbox{,} \\
   \label{CurvXi4b}
   \xi_{5}            &=& \frac{1}{\sqrt{2}} \left(\alpha_{13} -  \alpha_{12} \right)\hbox{,} \\
   \label{e:sin:brho}
   \xi_{6}            &=& \sin \rho_e - \sin {\bar \rho} \hbox{,}
\end{eqnarray}
where
\begin{equation}
\label{V0}
   \sin {\bar \rho} \, = \, \frac{2}{\sqrt{3}} \,\sin [ (\alpha_{23} + \alpha_{13} + \alpha_{12}) /6]
\end{equation}
$\sin\rho_e$ is the equilibrium value of $\sin\bar\rho$, $a$ is a molecular parameter, and $\Delta f_{ijk\ldots}$ are expansion coefficients. Here $r_i$ is a bond length and $\alpha_{ij}$ is an interbond bond angle.
Further details of this functional form and symmetry relations between $\Delta f_{ijk\ldots}$ can be found elsewhere \citep{05YuCaJe.NH3,jt554}.

The refined potential coefficients
$\Delta f_{ijk\ldots}$ were determined using a least squares fitting algorithm
which uses the derivatives of energies with respect to $\Delta f_{ijk\ldots}$ computed
via the Hellmann-Feynman theorem \citep{39Fexxxx.DMS}. The process
starts by setting all $\Delta f_{ijk\ldots} = 0$.
The resulting refined PES is only an `effective' one
since it depends on any approximation in the nuclear motion calculations; it is
therefore dependent
on the levels of kinetic energy (KE)  and PES expansion, and basis set used
(see
below).
As a result of this, improving the nuclear motion calculation may lead
to worse agreement with the observations.


The experimental data for the energies is taken from the extensive
high resolution infrared studies of Maki and co-workers
\citep{73KaMaDo.SO3,89OrEsMa.SO3,01ChVuMa.SO3,01MaBlSa.SO3,02BaChMa.SO3,
  03ShBlSa.SO3,04MaBlSa.SO3}.  The majority of these studies provide
upper and lower energy states labelled by their vibrational normal
mode and rotational ($J$, $K$) quantum numbers, which were validated
using effective Hamiltonians. However, the bands studied by
\citet{04MaBlSa.SO3} label transitions by rotational and vibrational
quantum numbers, but do not list upper and lower energy levels.
Combination differences were used to obtain energies for these bands
using the experimental line positions reported by the accompanying
publications; these are highlighted in Table \ref{rmsBands_hot}.  In
matching experimental and computed energies, a number of
experimentally derived energies were not included in the fit; these
correspond to transitions excluded by Maki \etal\ from their 
Hamiltonian fits.  A total of 119 energy levels for $J \leq$ 5 were
chosen from this set based on their reliability at reproducing the
observed transitions, with the condition that they are physically
accessible states with $A\p$ or $A\pp$ symmetry; any published values
of experimentally-derived purely vibrational terms (i.e. band centres)
that are inaccessible were not included. Table \ref{so3_refinement_j5}
lists all the $J$ = 5 levels used in the refinement process, comparing
with their final computed counterparts.

\begin{table}
\caption{Comparison of RMS differences for the \ai\ (UYT) and
refined (UYT2) PES for observed vibrational band centres of \sothree.
Values for band centres are the experimental ones of
\citep{01MaBlSa.SO3}. All values are in cm$^{-1}$.
Note: UYT does not cover transition frequencies above 4000 \cm. }
\centering
\begin{tabular}{lrrr}
\hline\hline
Band & Band Centre & UYT & UYT2\\
\hline

2$\nu_2$ - $\nu_2$	&	497.45	&	0.73	&	0.09	\\
$\nu_2$ - $\nu_0$	&	497.57	&	0.77	&	0.05	\\
$\nu_2$ + $\nu_4$ - $\nu_4$	&	497.81	&	0.82	&	0.03	
\\
2$\nu_4^{(l_4 = 0)}$ - $\nu_4$	&	529.72	&	1.33	&	0.30	
\\
$\nu_4$ - $\nu_0$	&	530.09	&	1.41	&	0.09	\\
$\nu_2$ + $\nu_4$ - $\nu_2$	&	530.33	&	0.22	&	0.08	
\\
2$\nu_4^{(l_4 = 2)}$ - $\nu_4$	&	530.36	&	1.54	&	0.37	
\\
$\nu_1$ - $\nu_4$	&	534.83	&	0.47	&	0.20	\\
$\nu_3$ - $\nu_0$	&	1391.52	&	4.06	&	0.09	\\
2$\nu_2$ + $\nu_4$ - $\nu_0$	&	1525.61	&	0.19	&	0.08	
\\
$^{\S}$$\nu_2$ + 2$\nu_4^{(l_4 = 0)}$ - $\nu_0$	&	1557.88	&	2.39	
&	1.17	\\
$^{\S}$$\nu_2$ + 2$\nu_4^{(l_4 = 2)}$ - $\nu_0$	&	1558.52	&	2.12	
&	0.64	\\
$^{\S}$$\nu_1$ + $\nu_2$ - $\nu_0$	&	1560.60	&	1.14	&	
1.28	\\
$^{\S}$3$\nu_4^{l_4 = 1}$ - $\nu_0$	&	1589.81	&	6.73	&	
4.00	\\
$^{\S}$$\nu_1$ + $\nu_4^{(l_4 = 1)}$ - $\nu_0$	&	1593.69	&	3.32	
&	3.57	\\
$^{\star}$$^{\S}$($\nu_3$ + $\nu_4$)$^{(L = 2)}$ - $\nu_0$	&	1917.68	
&	5.34	&	0.65	\\
2$\nu_3^{(l_3 = 2)} - \nu_0$	&	2777.87	&	7.53	&	0.20	
\\
$^{\S}$3$\nu_3^{(l_3 = 1)}$ - $\nu_0$	&	4136.39	&	--	&	
0.08	\\

\hline\hline
\end{tabular}

\mbox{}\\
{\flushleft
$^{\star}$The value $L$ is given by $L$ = $|l_3 + l_4|$, see
\citet{04MaBlSa.SO3}.\\
$^{\S}$These bands are refined using energy levels obtained from the data by \citep{01MaBlSa.SO3,04MaBlSa.SO3} via
combination differences.\\
}
\label{rmsBands_hot}
\end{table}

\begin{table}
\caption{Comparisons of vibrational ($J$ = 0) terms for \sothree,
between experimental  values \citep{01MaBlSa.SO3}, and
the
\ai\ (UYT) and refined (UYT2) PES. All values are given in \cm.}
\centering
\begin{tabular}{lrrr}
\hline\hline
& Obs.& UYT & UYT2\\
\hline
$\nu_2$	&	497.57	&	498.48	&	497.56	\\
$\nu_4$	&	530.09	&	528.59	&	530.09	\\
2$\nu_2$	&	995.02	&	995.35	&	993.67	\\
$\nu_2$ + $\nu_4^{(l_4 = 1)}$	&	1027.90	&	1027.35	&	1027.33	
\\
2$\nu_4^{(l_4 = 0)}$	&	1059.81	&	1056.50	&	1059.48	\\
2$\nu_4^{(l_4 = 2)}$	&	1060.45	&	1057.38	&	1060.45	\\
$\nu_1$	&	1064.92	&	1065.75	&	1066.49	\\
$\nu_3$	&	1391.52	&	1387.45	&	1391.51	\\
3$\nu_2$	&	1492.35	&	1490.76	&	1488.47	\\
2$\nu_2$ + $\nu_4^{(l_4 = 1)}$	&	1525.61	&	1524.48	&	1524.20	
\\
$\nu_2$ + 2$\nu_4^{(l_4 = 0)}$	&	1557.88	&	1555.59	&	1557.50	
\\
$\nu_2$ + 2$\nu_4^{(l_4 = 2)}$	&	1558.52	&	1556.45	&	1558.46	
\\
$\nu_1$ + $\nu_2$	&	1560.60	&	1565.33	&	1565.07	\\
3$\nu_4^{(l_4 = 1)}$	&	1589.81	&	1586.46	&	1588.97	\\
3$\nu_4^{(l_4 = 3)}$	&	1591.10	&	1586.43	&	1591.06	\\
$\nu_1$ + $\nu_4^{(l_4 = 1)}$	&	1593.69	&	1593.36	&	1595.92	
\\
$\nu_2$ + $\nu_3^{(l_3 = 1)}$	&	1884.57	&	1881.53	&	1884.29	
\\
$^{\star}$($\nu_3$ + $\nu_4$)$^{(L = 2)}$	&	1917.68	&	1912.24	
&	1917.68	\\
$^{\star}$($\nu_3$ + $\nu_4$)$^{(L = 0)}$	&	1918.23	&	1914.56	
&	1919.63	\\
2$\nu_3^{(l_3 = 0)}$	&	2766.40	&	2759.12	&	2766.38	\\
2$\nu_3^{(l_3 = 2)}$	&	2777.87	&	2770.29	&	2777.86	\\
3$\nu_3^{(l_3 = 1)}$	&	4136.39	&	4126.78	&	4136.33	\\
\hline\hline

\end{tabular}

\mbox{}\\
{\flushleft
$^{\star}$The value $L$ is given by $L$ = $|l_3 + l_4|$, see
\citet{04MaBlSa.SO3}.\\
}

\label{fundamentals_so3hot}
\end{table}

\begin{table}
\caption{Observed \citep{01MaBlSa.SO3}  minus Calculated residuals for the $J$ =
5 energy levels used
in the refinement procedure. All values are in \cm. The corresponding values
for
$J \le $ 5 are given in \citet{Underwood.thesis}.}
\centering
\begin{tabular}{llrrr}
\hline\hline
State & $K$ & Obs. & UYT2 & Obs. - Calc.\\
\hline
$\nu_0$	&	3	&	8.885	&	8.886	&	-0.001	\\
$\nu_2$	&	3	&	506.367	&	506.360	&	0.008	\\
	&	0	&	507.900	&	507.893	&	0.007	\\
	&	5	&	535.323	&	535.312	&	0.011	\\
$\nu_4^{(l_4 = 1)}$	&	4	&	538.471	&	538.490	&	
-0.020	\\
	&	2	&	539.561	&	539.560	&	0.001	\\
	&	1	&	540.677	&	540.685	&	-0.008	\\
2$\nu_2$	&	3	&	1002.357	&	1002.411	
&	-0.054	\\
$\nu_2$ + $\nu_4^{(l_4 = 1)}$	&	5	&	1033.102	&	
1033.058	&	0.044	\\
	&	4	&	1036.241	&	1036.226	&	
0.016	\\
	&	2	&	1037.252	&	1037.219	&	
0.033	\\
	&	1	&	1038.243	&	1038.220	&	
0.023	\\
	&	5	&	1068.278	&	1068.303	&	
-0.025	\\
2$\nu_4^{(l_4 = 0)}$	&	3	&	1068.461	&	
1068.456	&	0.005	\\
2$\nu_4^{(l_4 = 2)}$	&	2	&	1071.024	&	
1071.031	&	-0.007	\\
	&	5	&	1398.427	&	1398.437	&	
-0.010	\\
$\nu_3^{(l_3 = 1)}$	&	2	&	1401.580	&	
1401.581	&	-0.001	\\
	&	1	&	1401.599	&	1401.591	&	
0.009	\\
	&	5	&	1529.365	&	1529.362	&	
0.003	\\
2$\nu_2$ + $\nu_4^{(l_4 = 1)}$	&	4	&	1532.498	&	
1532.520	&	-0.022	\\
	&	2	&	1533.442	&	1533.448	&	
-0.006	\\
$\nu_1$ + $\nu_2$	&	3	&	1573.870	&	
1573.856	&	0.014	\\
	&	0	&	1575.400	&	1575.387	&	
0.013	\\
3$\nu_4^{(l_4 = 1)}$	&	4	&	1597.408	&	
1597.410	&	-0.002	\\
$\nu_1$ + $\nu_4^{(l_4 = 1)}$	&	5	&	1601.162	&	
1601.150	&	0.012	\\
	&	4	&	1604.308	&	1604.322	&	
-0.014	\\
	&	2	&	1605.430	&	1605.426	&	
0.004	\\
	&	1	&	1606.574	&	1606.577	&	
-0.002	\\
	&	5	&	1923.797	&	1923.808	&	
-0.011	\\
$^{\star}$($\nu_3$ + $\nu_4$)$^{(L = 2)}$	&	4	&	
1925.310	&	1925.318	&	-0.008	\\
	&	0	&	1927.488	&	1927.422	&	
0.066	\\
	&	1	&	1927.982	&	1927.988	&	
-0.006	\\
	&	5	&	2782.262	&	2782.227	&	
0.035	\\
2$\nu_3^{(l_3 = 2)}$	&	4	&	2786.812	&	
2786.837	&	-0.025	\\
	&	2	&	2786.901	&	2786.888	&	
0.014	\\
	&	1	&	2788.419	&	2788.425	&	
-0.006	\\
3$\nu_3^{(l_3 = 1)}$	&	5	&	4143.316	&	
4143.246	&	0.070	\\
	&	2	&	4146.379	&	4146.329	&	
0.050	\\
\hline\hline

\end{tabular}

$^{\star}$The value $L$ is given by $L$ = $|l_3 + l_4|$, see
\citet{04MaBlSa.SO3}.

\label{so3_refinement_j5}
\end{table}

Table \ref{rmsBands_hot} shows the effect of the final potential refinement on
the bands used in the refining procedure. The root mean square (RMS)
differences are calculated by matching all experimental lines for each band
with calculated values via their quantum number assignments for all  $J\le$ 5 available.

The RMS differences calculated are slightly increased when including
higher $J >5$ term values comparing to the residuals in
Table~\ref{so3_refinement_j5} for a few bands as a result of the
refinement, namely the $\nu_1$ + $\nu_2$ and $\nu_1$ + $\nu_4^{(l_4 =
  1)}$ bands. The experimental energy levels used to refine these two
bands were obtained using combination differences. However, for some
rotationally excited levels within these bands, the quantum number
labelling of the experimental transitions appears dubious:
in particular there are a
number of transition whose labels are duplicated. These transitions were
not included in the RMS difference calculations but there must be some doubt
about the validity of the quantum number assignments of
the other transitions in these bands. This may well explain
the increased the RMS difference.

Table \ref{fundamentals_so3hot} compares all  published  vibrational ($J$ = 0)
term values with those calculated with \TROVE\ before and after refinement.
There are some discrepancies introduced by the refinement
procedure and in some cases deteriorations from the pre-refined values (e.g
2$\nu_2$). The quality of the refinement can be assessed from Table \ref{rmsBands_hot}, with the exception of 3$\nu_2$, for which there is no experimental band data available
beyond the quoted vibrational term value \citep{04MaBlSa.SO3}.

Our refined PES is given as Supplementary Information to this article.

\subsection{Calculation using \TROVE}
\label{s:calcTroveHot}

In specifying a calculation using \TROVE, it is necessary to
fix a number of parameters. In particular both the KE and PES are
expanded as a Taylor series about the equilibrium geometry \citep{07YuThJe.method}.
For UYT the expansions were truncated at 4$^{\rm th}$ and 8$^{\rm th}$ orders respectively. Here the KE expansion order was  increased to 6$^{\rm th}$ in order to allow better convergence. For the detailed description of the basis set see \citet{jt554}. Here it suffices to define the maximal polyad number $P_{\rm max}$ used
in \TROVE\ to control the size of the basis set.
The polyad number in case of \sothree\ in terms of the normal mode quantum numbers is given by
\begin{equation}
\label{e:polyad}
P = 2(n_1 + n_3) + n_2 + n_4,
\end{equation}
where $n_1, n_2, n_3$ and $n_4$ are the normal mode quanta associated with the $\nu_1, \nu_2, \nu_3$ and $\nu_4$ vibrational modes. For the UYT2 calculations the value of $P_{\rm max}$  was set initially to $24$ for the 1D primitive basis functions to form a product-type basis set, which was then contracted to $P_{\rm max}  = 18$ after a set of prediagonalizations of reduced Hamiltonian matrices and symmetrized.   The value of $P_{\rm max}$ used for UYT was 12. This increase was necessary to allow both better convergence of the increased number of energies which is needed for high temperature spectra. Only energies lying up to 10~000~\cm\ above the ground state were considered as part of this study.

The high symmetry of $^{32}$S$^{16}$O$_{3}$, and the associated nuclear spin statistics,
means that it is only necessary to consider
transitions between $A_1\p$ and $A_1\pp$ symmetries of the \Dh\
point group used for the calculations. The final UYT2 line list
consists of all allowed transitions between 0 $ < \nu \leq$ 5000 \cm,
satisfying
the conditions $E' \leq 9000$ \cm, $E''  \leq  4000$ \cm\ and
$J \leq$ 130. These parameters are designed to give a complete spectrum up to
5000 \cm\ ($\lambda > 2$ $\mu$m) for temperatures up to about 800 K.
Generating a complete line list with these parameters is computationally
demanding and therefore requires special measures to be taken.

In terms of memory,  diagonalisation of the Hamiltonian matrices
is the most computationally expensive part of the line list calculation. For
each $J$, a matrix is built
\citep{07YuThJe.method,jt466}  and then stored
in the memory for diagonalisation, using an appropriate eigensolver routine.
\TROVE\ uses a symmetry adapted basis set representation and allows splitting of each
$J$ Hamiltonian matrix further into the six symmetry blocks ($A_1\p$, $A_2\p$, $E\p$, $A_1\pp$, $A_2\pp$, and $E\pp$) which are dealt with separately. Since only the $A_1\p$ and $A_1\pp$ symmetry species are allowed by the nuclear statistics of
$^{32}$S$^{16}$O$_{3}$, only these symmetry blocks are diagonalized.

Memory requirements scale with the square of the dimension of the Hamiltonian
matrix, $N^{\rm Ham}_J$. This is given roughly by  $N^{\rm Ham}_J = N_{J = 0} \times (2J + 1)$,
where $N_{J = 0}$ is the dimension of the purely vibrational matrix. For UYT2,
the combined dimension $N_{J = 0}$ of both $A_1\p$ and $A_1\pp$ symmetries is
2692; for comparison, the UYT line list calculations used $N_{J
= 0}$ = 679. The size of the largest matrix considered in the room-temperature
calculations (for $J$ = 85) is $N^{\rm Ham}_{85} = 111~296$, which is already
surpassed by $J$ = 21 for UYT2 for which the value of $N^{\rm Ham}_{85}
= 454~488$. It became quickly apparent that the diagonalisation techniques
previously employed to determine the UYT wavefunctions would be impractical
for UYT2.

Nuclear motion calculations were performed using
both the Darwin and COSMOS HPC facilities in Cambridge, UK.
Each of the computing nodes on the Darwin cluster provide 16 CPUs across two
2.60 GHz 8-core Intel Sandy Bridge E5-2670 processors, and a maximum of 64 Gb
of RAM. The advantage of moving eigenfunction calculations to the Darwin
cluster are that an entire node can be dedicated to one calculation, spread
across the 16 CPUs. Since
multiple nodes can be accessed by a single user at any time, multiple
computations were carried out simultaneously.

Diagonalisation of  matrices with $J \leq$ 32 was possible using  the LAPACK
DSYEV eigensolver \citep{99AnBaBi.method}, optimised for OpenMP parallelisation across
multiple (16) CPUs.
For 32 $< J \leq$ 90 a distributed memory approach was used with an MPI-optimised
version of the eigensolver, PDSYEVD, which allowed diagonalisations across
multiple Darwin/COSMOS nodes in order to make use of their collective memory.
In order to diagonalise the matrix within the 36 hour wall clock limit, it was
necessary to perform this method in three steps. First, for a given $J$ and
symmetry species $\Gamma$, the Hamiltonian matrix was constructed and saved to
disk. Secondly, the matrix was then read and diagonalised using PDSYEVD
across the number of nodes required to store of the matrix in their
shared memory. This produces a set of eigenvectors which were
read in again to convert into the \TROVE\ eigenfunction format.

For $J >$ 90 yet another approach was developed for use on the COSMOS shared
memory machine. This method employed the PLASMA DSYTRDX routine \citep{13KuLuYa.method}
and, unlike the above procedure, constructed, diagonalised and stored
wavefunctions to disk in a single process by extending both the standard wallclock time
and memory limits.
For $J$ = 130 ($\Gamma$ = $A_1\p$) a total of 52 hours of
real time was taken to construct and diagonalize the Hamiltonian matrix across
416 CPUs, and utilising 3140 Gb of RAM.

While matrix diagonalisation dominates the memory requirements of
the calculation, computing the line strengths, $S(f \leftarrow i)$,
is the major user of computer time.
In principle line strengths for all transitions obeying the rigorous
electric dipole selection rules, $\Delta J = J\p - J\pp = 0, \pm
1 \ (J\pp + J\p \ge 1)$ and $A_1\p
\leftrightarrow A_1\pp$, were computed.
In practice this was modified to reduce the
computational demands. Firstly, calculations of the line strength  only take
into consideration the basis functions of the final state wavefunction whose
coefficients are greater than 10$^{-14}$. In addition to this a threshold value
for the Einstein A coefficient of 10$^{-74}$ s$^{-1}$ dictates which
transitions
are kept. However, the number of line strength calculations to be performed
still remains very large
and  even with parallelisation  across multiple
Darwin CPUs, performing the calculations proved to be both computationally
expensive and difficult.

To help expedite these computations, an adapted version of \TROVE\ was used
which is optimised for performing calculations on graphical processing units (GPUs). The use of
this implementation, known as GAIN (GPU Accelarated INtensities)
\citep{jtGAIN}, allowed for the computation
of transition strengths for the more computationally demanding parts of the
calculations. These calculations were performed on the Emerald GPU cluster,
based  in Southampton. In general, the calculation of transition strengths
across multiple GPUs was much faster than the Darwin CPUs. For example, there
are a total of 349~481~979 transitions for $J\pp$ = 35, which took a total of
17~338 CPU hours to compute on the Darwin nodes, compared to 2053 GPU hours on
the Emerald nodes for 346~620~894 transitions for the larger $J\pp$ = 59 case.
These GPU calculations were carried out for those $J \leftrightarrow\ J + 1$
pairs containing a large number of states, while the Darwin CPUs were reserved
for the less computationally demanding sections.

21 billion transitions were calculated for UYT2,
which is two orders of magnitude larger than UYT.
Overall performing the computations
needed for the UYT2 line list took us over 2 years.

\section{Experiments}
\begin{landscape}
\begin{figure}
\includegraphics[width=20cm]{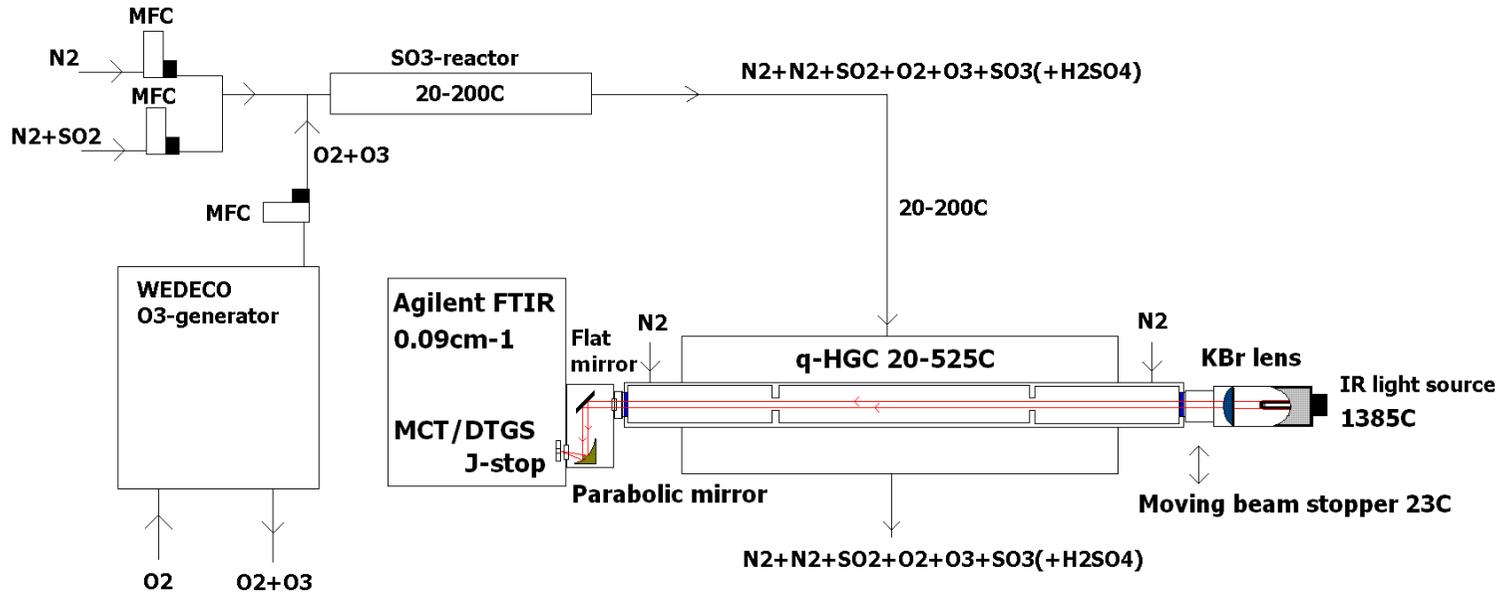}
\caption{Set up used for SO$_2$/SO$_3$/O$_3$ infrared absorption measurements.}
\label{f:expt}
\end{figure}
\end{landscape}

SO$_3$ absorbance measurements at temperatures up to 500~C were
performed using a quartz high-temperature gas flow cell (q-HGC). The
cell has been described in details by \citet{13GrFaNi.SO2} and has
recently been used for measurements with NH$_3$ \citep{jt616},
S-containing gases \citep{14GrFaCl.SO3} and some PAH compounds \citep{15GrSoFa.PAH}.

Because SO$_3$ is an extremely reactive gas and normally contains traces of
SO$_2$ if ordered from a gas supplier, it was decided to produce SO$_3$
directly in the set up. It is known that SO$_2$ can react with O$_3$ and
form SO$_3$:
\begin{equation}
{\rm SO}_2 + {\rm O}_3 \rightarrow {\rm SO}_3  + {\rm O}_2. \label{r1}
\end{equation}
 The rate constant for the reaction~(\ref{r1}) is
temperature dependent: higher temperatures favour SO$_3$ formation.
However at higher temperatures O$_3$ starts to decompose into O$_2$ and O:
\begin{equation}
{\rm O}_3 (+ {\rm M})  \rightarrow {\rm O}_2 + {\rm O} (+ {\rm M})  \label{r2}
\end{equation}
Some O and O$_2$ can contribute further in
SO$_3$ formation and ``re-cycle'' O$_3$:
\begin{eqnarray}
\nonumber
{\rm SO}_2 + {\rm O} (+ {\rm M})  &\rightarrow& {\rm SO}_3 (+ M) \\
\nonumber
{\rm O} + {\rm O}_2 (+{\rm M})  &\rightarrow& {\rm O}_3 (+ {\rm M}) \\
{\rm O} + {\rm O} (+ {\rm M})  &\rightarrow& {\rm O}_2 (+ {\rm M})
\end{eqnarray}
If any water traces  are present
in the system, SO$_3$ will rapidly be converted into sulfuric acid
(H$_2$SO$_4$):
\begin{equation}
{\rm SO}_3 + {\rm H}_2{\rm O} (+ {\rm M}) \rightarrow  {\rm H}_2{\rm SO}_4 (+ {\rm M}),
\end{equation}
which is then followed by further surface-promoted reaction:
\begin{equation}
{\rm H}_2{\rm SO}_4 + {\rm surface}  \rightarrow {\rm products}.
\end{equation}
Other possible SO$_3$ removal channels
are:
\begin{eqnarray}
\nonumber
{\rm SO}_3 (+ {\rm M})  &\rightarrow& {\rm SO}_2 + {\rm O} (+ {\rm M}) \\
{\rm SO}_3 + {\rm surface} &\rightarrow& {\rm products}.
\end{eqnarray}
These reactions are very prominent in a clean set up with ``active'' surfaces.

In the presence of O$_2$, a reversible reaction (which is also
temperature-dependent) takes place:
\begin{equation}
2 {\rm SO}_2 + {\rm O}_2 (+ {\rm M})  \rightarrow 2 {\rm SO}_3 (+ {\rm M}).
\label{r3}
\end{equation}
However SO$_3$ formation through reaction~(\ref{r3}) takes place at
temperatures higher than 500 C.  The set-up is shown in
Fig.~\ref{f:expt}. It can be divided into two parts: a SO$_3$ generation
part and a part for optical measurements.

The optical
part of the set up includes a high-resolution FTIR spectrometer
(Agilent 660 with a linearized broad-band MCT detector), the q-HGC and
a light-source (Hawkeye, IR-Si217, 1385C) with a KBr plano-convex
lens. The light source is placed in the focus of the KBr lens. The
FTIR and sections between the FTIR/q-GHC and q-HGC/IR light source
have been purged by CO$_2$/H$_2$O-free air obtained from a purge generator.

The O$_3$-generation part consists of a set of high-end mass-flow
controllers (MFC's), O$_3$-generator and a unit called SO$_3$-reactor.
MFC's (Bronkhorst) have been used to keep constant gas flows and mix
gas flows of N$_2$, N$_2$+SO$_2$ and O$_2$+O$_3$ in desirable ratios.
An O$_3$-generator (WEDECO GSO 30, water cooled, rated capacity at
full load 100 g hour$^{-1}$ of O$_3$ with use of O$_2$) was used to
produce O$_3$ from O$_2$. Because of the high O$_2$ flow rate required
for stable operation of the O$_3$-generator, only a part of the
O$_2$+O$_3$ flow was used in the measurements. The ozone generator was
operated at about 30\% (225 W) of the full load.  SO$_3$-reactor was a
50 cm heated quartz tube (20 -- 200 C) with inner diameter of 50 mm.
N$_2$+SO$_2$ was mixed in the SO$_3$-reactor with O$_2$+O$_3$ from the
ozone-generator at 170 -- 190 C. The gas residence time in the
O$_3$-reactor was about 60 s at 1 l$_n$ min$^{-1}$ (normal litres per
minute) flow rate which was enough to convert about 50\% of SO$_2$
into SO$_3$ and at the same time mostly decompose O$_3$. The
SO$_3$-reactor was connected through a heated Teflon-line (inner
diameter 4 mm, $T=$ 20 -- 200 C) to the inlet of the q-HGC. The gas
residence time in the Teflon-line was about 0.8 s (at 1 ln/min) and
some further (minor) conversion of SO$_2$ to SO$_3$ took also place.

Bottles with premixed gas mixture, N$_2$ + SO$_2$ (5000ppm)
(Strandm\"{o}llen) and N$_2$/O$_2$ (99.998\%) (AGA) have been used for
reference and SO$_2$/SO$_3$ absorbance measurements. The main flow in
the q-HGC was balanced with the two buffer flows of N$_2$ from q-HGC's
buffer parts. Most of SO$_2$/SO$_3$ absorbance measurements have
performed at 0.25-0.5 \cm\ nominal spectral resolutions and around
atmospheric pressure in the q-HGC. Few measurements have been
performed at 0.09~\cm\ spectra resolution.  The measurements were
performed in following steps:
\begin{enumerate}
\item N$_2$ + O$_2$ in q-HGC, reference spectra, ozone generator ``off'';
\item N$_2$ + O$_2$ + SO$_2$ (2500 ppm) in q-HGC, absorption spectra, ozone generator ``off'';
\item N$_2$ + O$_2$ + SO$_2$ + SO$_3$ + O$_3$ in q-HGC, absorption spectra, ozone generator ``on'', initial SO$_2$ concentration 2500~ppm;
\item N$_2$ + O$_2$ + O$_3$, in q-HGC, ozone generator ``on'' in order to measure O$_3$ traces in the q-HGC (addition step used only for some measurements).
\end{enumerate}
O$_3$ has several absorption bands in 400-6000 \cm, which do not
interfere with SO$_2$/SO$_3$ absorption bands. At each step two
measurements were made: with a light source (emission from the cell
and light source) and without a light source (emission from the cell).
Experimental absorption spectra SO$_2$/SO$_3$ were reconstructed in
the way described in section 3.1 of \citet{jt616}. Spectra of SO$_2$
measured in step~2 have been normalized and subtracted from the
composite SO$_2$+SO$_3$ spectra obtained in step~3 in order to get the
zero absorption signal in vicinity of the SO$_2$ bands as one can see
in Fig. \ref{f:SO2+SO3+O3}. It was further assumed that all SO$_2$ was
consumed to produce SO$_3$ (i.e. no SO$_3$ losses channels). Note
various log10-absorption scales on these figures.  The extra (weak)
broad feature in the region 1200-1285 \cm\ is caused by the O$_3$
production in the O$_3$-generator.

Figure~\ref{f:SO2+SO3:PNNL} gives a comparison of our newly measured cross
sections in the 7.4 $\mu$m region
with those available from the PNNL database for {SO$_2$ (upper) and SO$_3$ (lower).

\begin{figure}
\includegraphics[width=\columnwidth]{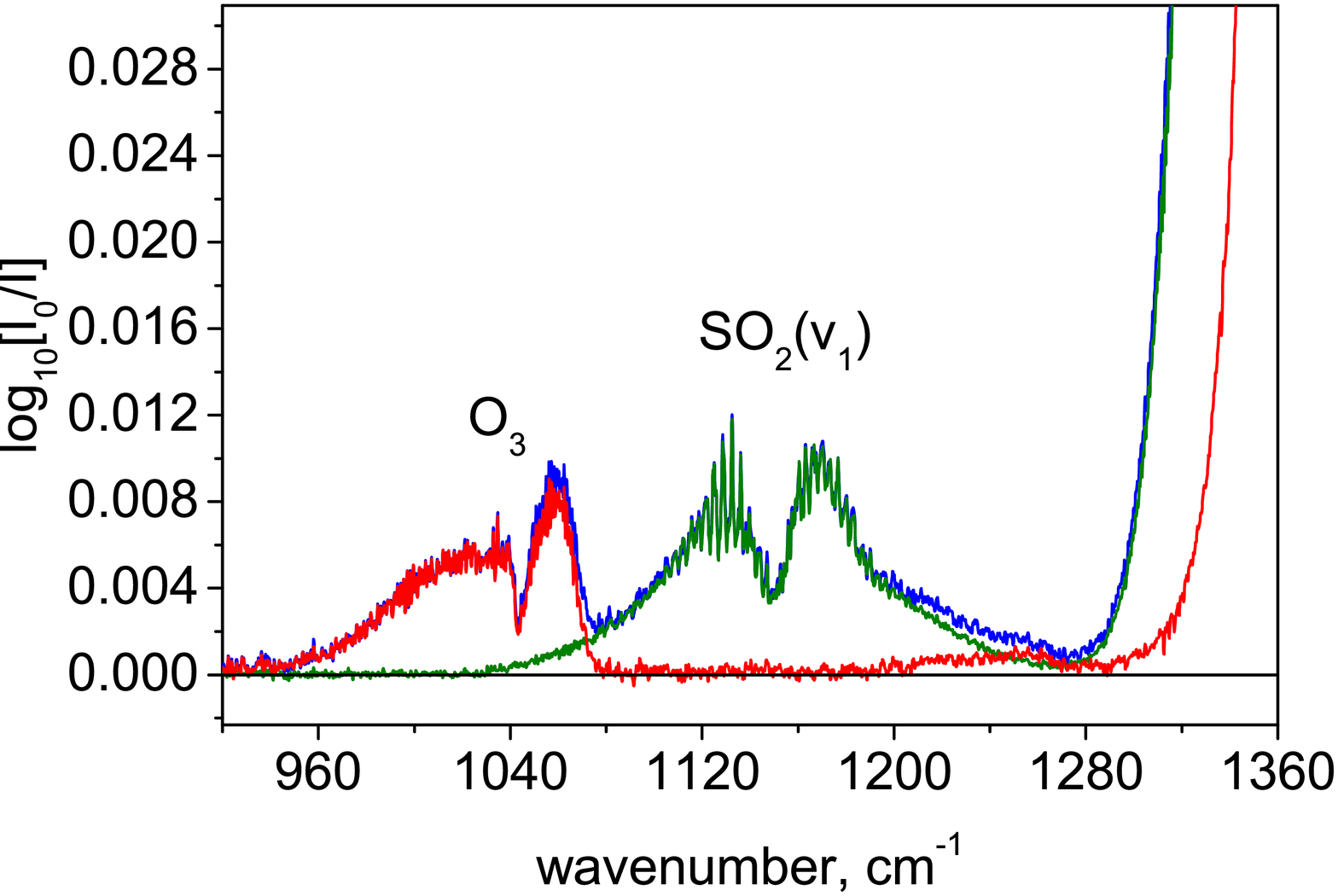} \\
\includegraphics[width=\columnwidth]{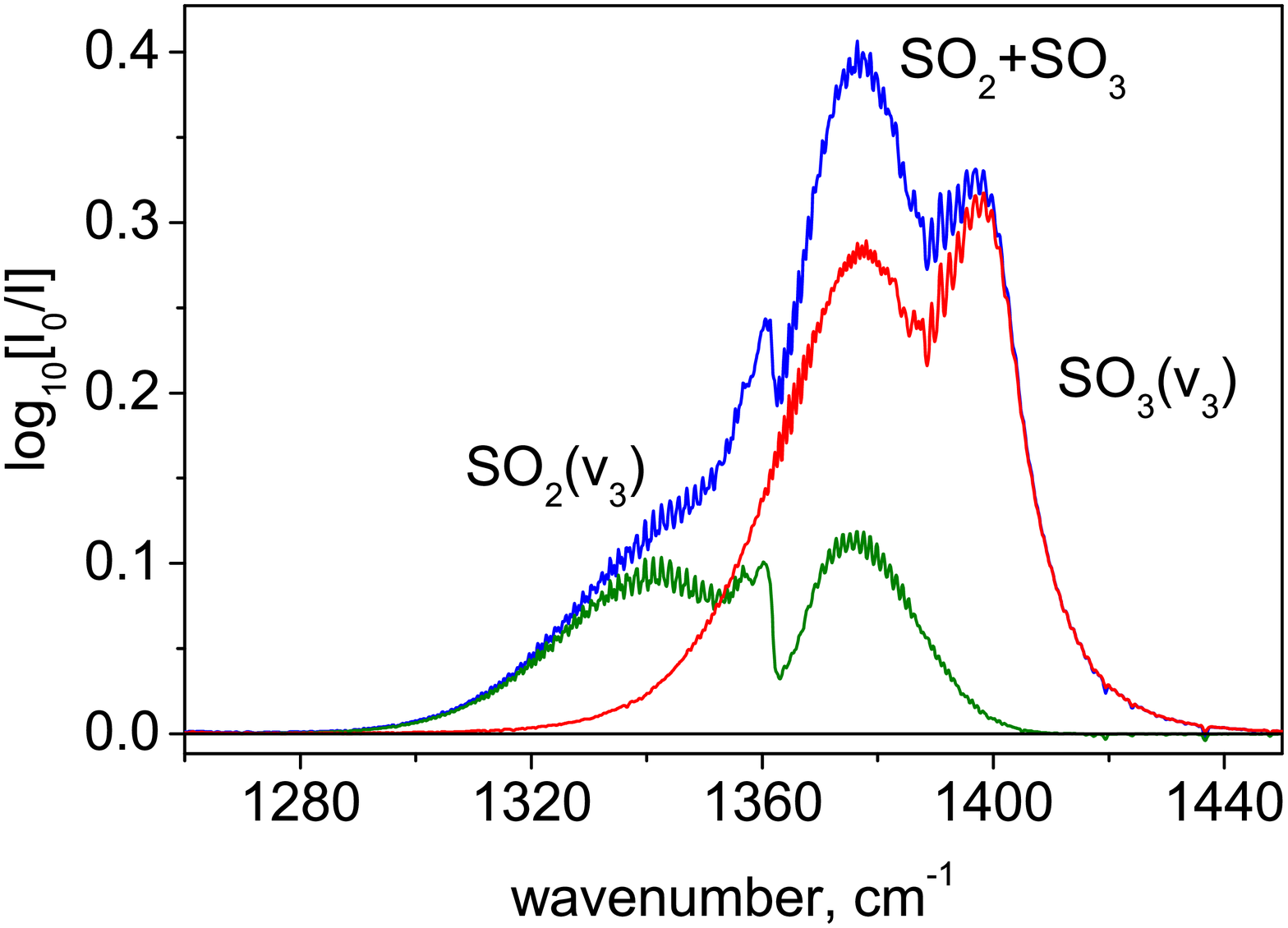} \\
\includegraphics[width=\columnwidth]{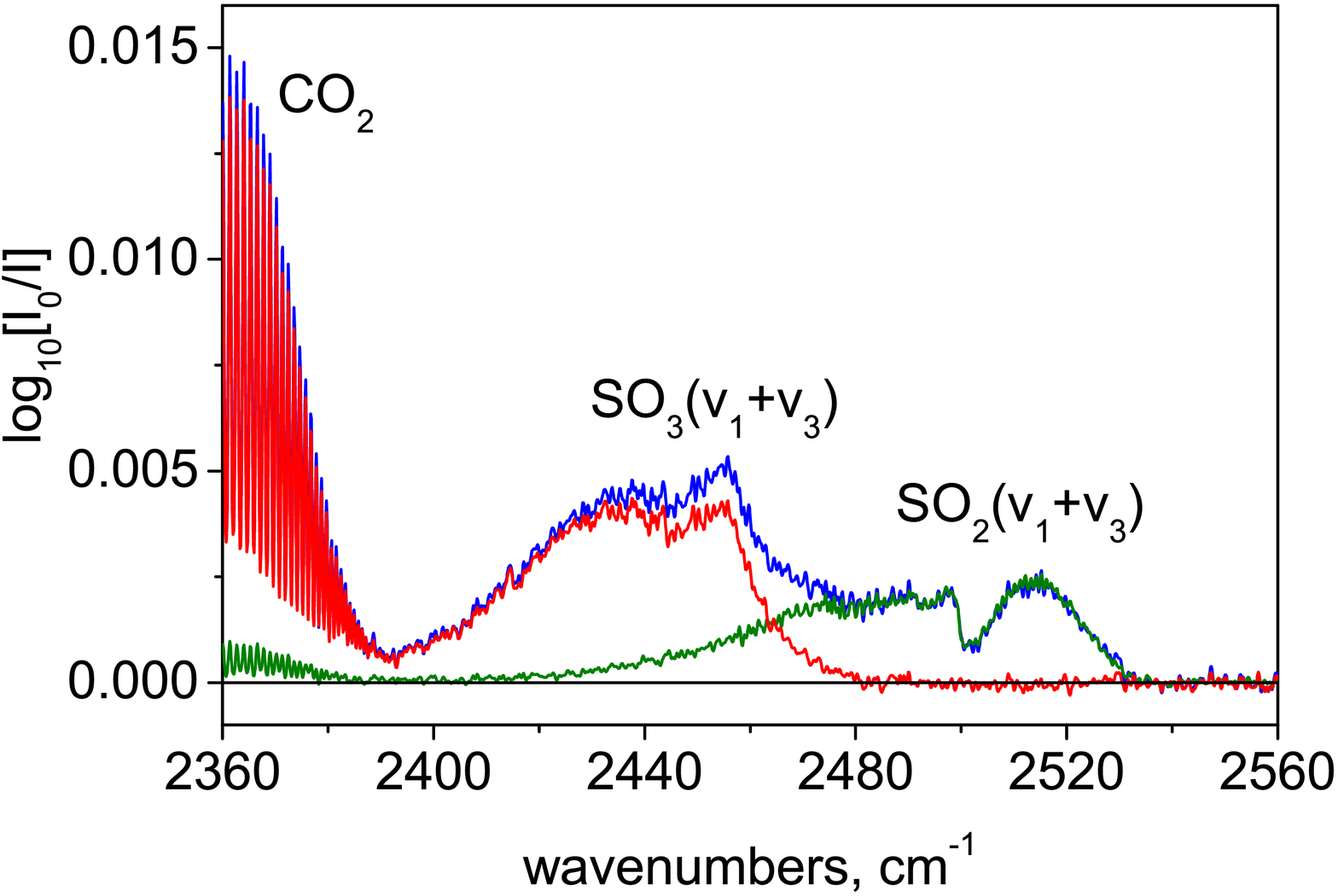}
\caption{Composite SO$_2$+SO$_3$+O$_3$ absorption spectrum at 300~C (blue) together with (normalized) SO$_2$ spectrum (olive) and the result of subtraction of SO$_2$ spectrum from the composite one (red): The vicinity of the $\nu_1$ (upper),  $\nu_3$ (middle) and $\nu_1+\nu_3$(lower) bands of SO$_2$.}
\label{f:SO2+SO3+O3}
\end{figure}

\begin{figure}
\includegraphics[width=\columnwidth]{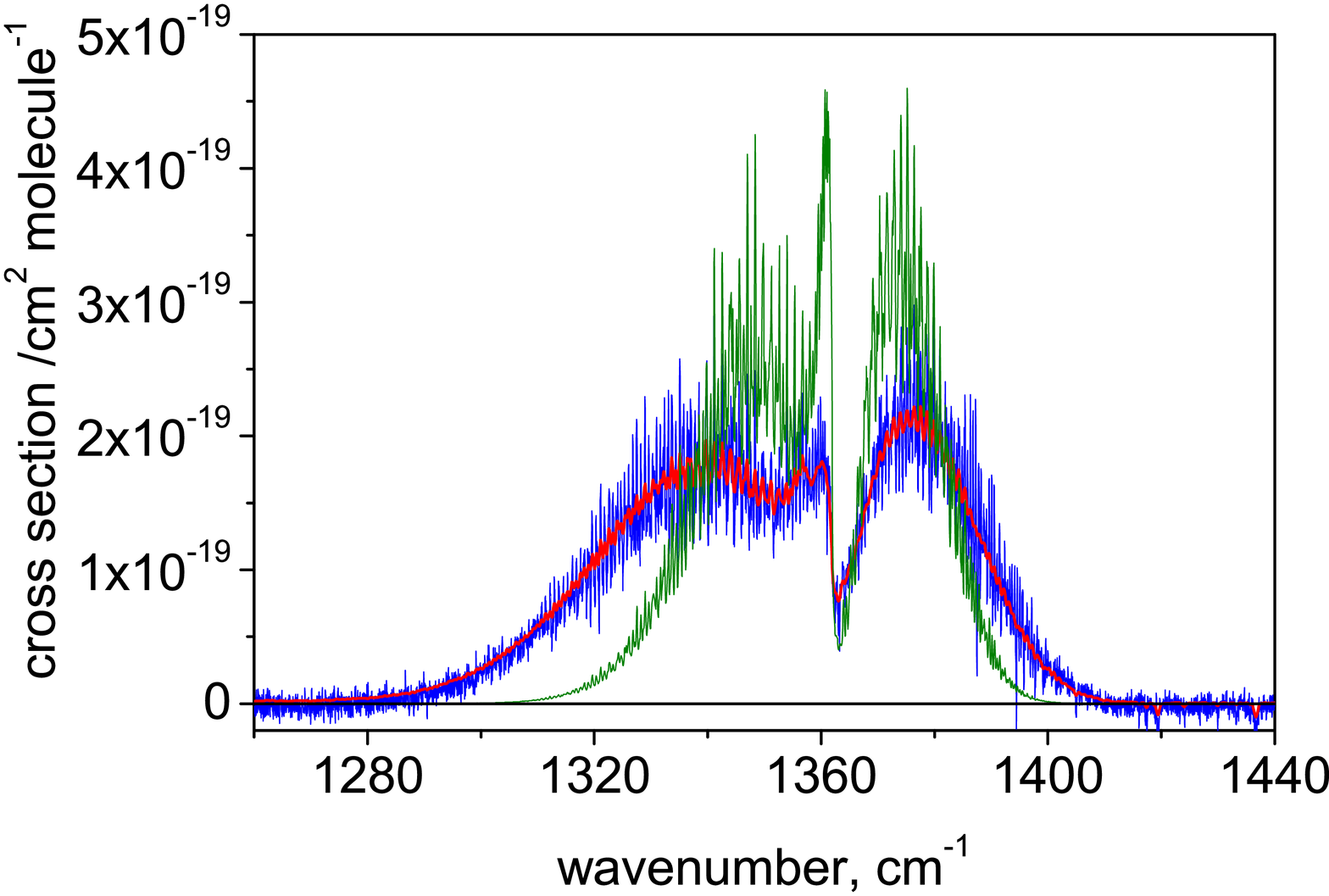} \\
\includegraphics[width=\columnwidth]{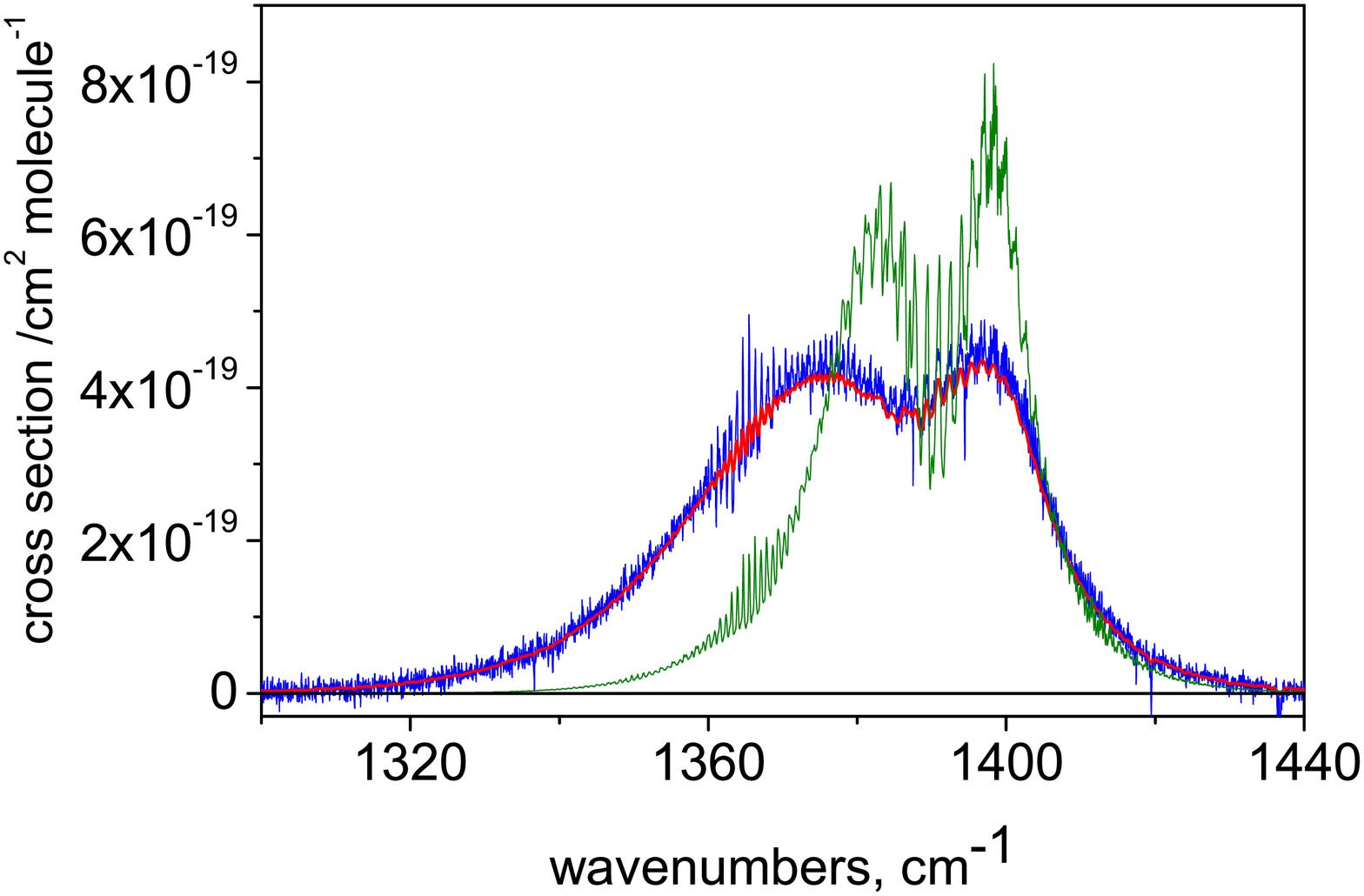}
\caption{SO$_2$ (upper) and SO$_3$ (lower) absorption cross sections at 400~C: 0.09 \cm\ (blue, boxcar), 0.5 \cm\ (red, triangular). Cross section from PNNL database at 25~C (olive, 0.112~\cm, boxcar) are shown for comparison.}
\label{f:SO2+SO3:PNNL}
\end{figure}

\section{Overview of the UYT2 Line List}

The UYT2 line list is presented in the ExoMol format \citep{jt548,jt631}
with main data contained in a states and a set of transtions files.
Tables~\ref{tab:states} and \ref{tab:trans} give portions of these
files. The complete files can be obtained
via
\url{ftp://cdsarc.u-strasbg.fr/pub/cats/J/MNRAS/xxx/yy}, or
\url{http://cdsarc.u-strasbg.fr/viz-bin/qcat?J/MNRAS//xxx/yy}, as well
as the exomol website, \url{www.exomol.com}.

\begin{table*}
\caption{ Extract from the UTY2 state file for SO$_3$;
quantum numbers are specified in Table~\ref{UYT2qn}. The full table is available from
http://cdsarc.u-strasbg.fr/cgi-bin/VizieR?-source=J/MNRAS/xxx/yy.}
\begin{tabular}{rrrrrrrrrrrrrrrrrrrr}
\hline\hline
$n$ & $\tilde{E}$& $g$ & $J$ & $\Gamma_{\rm Total}$ & $K$ & $\Gamma_{\rm Rot}$ & $v_1$
& $v_2$ & $v_3$ & $v_4$  & $v_5$ & $v_6$ & $n_1$ & $n_2$ & $n_3$ & $L_3$  & $n_4$ & $L_4$\ & $\Gamma_{\rm Vib}$\\
\hline
1	&	0.0000	&	1	&	0	&	1	&	0	&	1	&	0	&	0	&	0	&	0	&	0	&	0	&	0	&	0	&	0	&	0	&	0	&	0&	1	\\
2	&	993.6780	&	1	&	0	&	1	&	0	&	1	&	0	&	0	&	0	&	0	&	0	&	2	&	0	&	2	&	0	&	0	&	0	&	0&	1	\\
3	&	1059.4770	&	1	&	0	&	1	&	0	&	1	&	0	&	0	&	0	&	0	&	2	&	0	&	0	&	0	&	0	&	0	&	2	&	0&	1	\\
4	&	1066.4970	&	1	&	0	&	1	&	0	&	1	&	1	&	0	&	0	&	0	&	0	&	0	&	1	&	0	&	0	&	0	&	0	&	0&	1	\\
5	&	1591.0349	&	1	&	0	&	1	&	0	&	1	&	0	&	0	&	0	&	0	&	3	&	0	&	0	&	0	&	0	&	0	&	3	&	1&	1	\\
6	&	1919.6346	&	1	&	0	&	1	&	0	&	1	&	1	&	0	&	0	&	0	&	1	&	0	&	0	&	0	&	1	&	1	&	1	&	1&	1	\\
7	&	1981.9944	&	1	&	0	&	1	&	0	&	1	&	0	&	0	&	0	&	0	&	0	&	4	&	0	&	4	&	0	&	0	&	0	&	0&	1	\\
8	&	2054.0505	&	1	&	0	&	1	&	0	&	1	&	0	&	0	&	0	&	0	&	2	&	2	&	0	&	2	&	0	&	0	&	2	&	0&	1	\\
9	&	2061.9334	&	1	&	0	&	1	&	0	&	1	&	1	&	0	&	0	&	0	&	0	&	2	&	1	&	2	&	0	&	0	&	0	&	0&	1	\\
10	&	2117.4659	&	1	&	0	&	1	&	0	&	1	&	0	&	0	&	0	&	0	&	4	&	0	&	0	&	0	&	0	&	0	&	4	&	0&	1	\\
11	&	2124.4973	&	1	&	0	&	1	&	0	&	1	&	1	&	0	&	0	&	0	&	2	&	0	&	1	&	0	&	0	&	0	&	2	&	0&	1	\\
12	&	2129.3331	&	1	&	0	&	1	&	0	&	1	&	0	&	1	&	1	&	0	&	0	&	0	&	2	&	0	&	0	&	0	&	0	&	0&	1	\\
13	&	2444.1614	&	1	&	0	&	1	&	0	&	1	&	1	&	0	&	0	&	0	&	2	&	0	&	0	&	0	&	1	&	1	&	2	&	2&	1	\\
14	&	2586.0493	&	1	&	0	&	1	&	0	&	1	&	0	&	0	&	0	&	0	&	3	&	2	&	0	&	2	&	0	&	0	&	3	&	1&	1	\\
15	&	2648.2382	&	1	&	0	&	1	&	0	&	1	&	0	&	0	&	0	&	0	&	5	&	0	&	0	&	0	&	0	&	0	&	5	&	1&	1	\\
16	&	2655.7551	&	1	&	0	&	1	&	0	&	1	&	1	&	0	&	0	&	0	&	3	&	0	&	1	&	0	&	0	&	0	&	3	&	1&	1	\\
17	&	2766.3812	&	1	&	0	&	1	&	0	&	1	&	0	&	2	&	0	&	0	&	0	&	0	&	0	&	0	&	2	&	0	&	0	&	0&	1	\\
18	&	2904.3481	&	1	&	0	&	1	&	0	&	1	&	1	&	0	&	0	&	0	&	1	&	2	&	0	&	2	&	1	&	1	&	1	&	1&	1	\\

\hline\hline
\end{tabular}
\label{tab:states}
\end{table*}

\begin{table*}
\centering
\caption{Quantum numbers used in labelling energy states.}
\begin{tabular}{ll}
\hline\hline
Quantum Number & Description\\
\hline
$n$ & Counting index\\
$\tilde{E}$ & Energy value (\cm)\\
$g$ & Total degeneracy of the state\\
$J$ & Angular momentum quantum number\\
$\Gamma_{\rm Total}$ & Total symmetry in \Dh(M): 1 = $A_1\p$,  4 = $A_1\pp$\\
$K$ & Projection of $J$ on to the $z$-axis\\
$\Gamma_{\rm Rot}$ & Rotational symmetry in \Dh(M): 1 = $A_1\p$, 2 = $A_2\p$, 3 = $E\p$ , 4 = $A_1\pp$, 5 = $A_2\pp$, 6 = $E\pp$\\
$v_i, i =1-6$ & Local mode vibrational quantum numbers \\
$n_1$, $n_2$, $n_3$, $n_4$ & Normal mode vibrational quantum numbers\\
$L_3$, $L_4$ & $L$ projections of the vibrational angular momenta\\
$\Gamma_{\rm Vib}$ & Vibrational symmetry in \Dh(M): 1 = $A_1\p$, 2 = $A_2\p$, 3 = $E\p$ , 4 = $A_1\pp$, 5 = $A_2\pp$, 6 = $E\pp$\\
\hline
\end{tabular}
\label{UYT2qn}
\end{table*}

\begin{table}
\caption{Extract from the UTY2 transitions file for SO$_3$.
The full table is available from
http://cdsarc.u-strasbg.fr/cgi-bin/VizieR?-source=J/MNRAS/xxx/yy.}
\begin{tabular}{rrr}
\hline\hline
\multicolumn{1}{c}{$f$} & \multicolumn{1}{c}{$i$} & \multicolumn{1}{c}{A} \\
\hline
      237007   &    249581 & \verb!  1.1253e-17 ! \\
      158430   &    148459 & \verb!  2.8358e-17 ! \\
      549592   &    568676 & \verb!  1.3725e-16 ! \\
      120670   &    112002 & \verb!  1.4546e-16 ! \\
     2080392   &   2117071 & \verb!  9.0696e-18 ! \\
      289088   &    302965 & \verb!  1.4938e-16 ! \\
      393104   &    377035 & \verb!  1.5764e-16 ! \\
       43637   &     49289 & \verb!  2.1375e-16 ! \\
      587986   &    607961 & \verb!  2.0370e-16 ! \\
      587868   &    647986 & \verb!  4.2068e-18 ! \\
     2007259   &   2043487 & \verb!  5.2490e-18 ! \\
      627725   &    648113 & \verb!  3.0673e-16 ! \\
\hline
\hline
\end{tabular}
\label{tab:trans}

\noindent
 $f$: Upper  state counting number;
$i$:  Lower  state counting number; $A_{fi}$:  Einstein-A
coefficient in s$^{-1}$.

\end{table}

The energy levels listed in the states file are labelled with the quantum
numbers summarised in Table \ref{UYT2qn} and are based on those recommended
by \citet{jt546} for ammonia with the simplification that one does
not need to consider inversion.
Only quantum numbers $J$, $g_{\rm Total}$,
$\Gamma_{\rm Total}$ and the counting index, $n$  are rigorously defined. The remaining quantum numbers
represent the largest contribution from rotational and vibrational components
of the wavefunction expansion associated with a given state.
\TROVE\ provides local mode quantum numbers associated with the basis set construction scheme used \citep{jt554}.
The normal mode vibrational quantum numbers, $n_1$, $n_2$, $n_3$ and $n_4$, and or their angular momentum projections $L_3$ = $|l_3|$ and $L_4$ = $|l_4|$ were obtained from the local mode quantum numbers via the correlation rules
\begin{eqnarray}
\label{e:TROVEqnMap1}
n_1 + n_3 &=& v_1 + v_2 + v_3, \\
n_2 + n_4 &=& v_4 + v_5 + v_6,
\end{eqnarray}
and
\begin{eqnarray}
\label{e:TROVEqnMap2}
l_3 &=& -v_3, -v_3 + 2, ..., v_3 - 2,  v_3, \\
l_4 &=& -v_4, -v_4 + 2, ..., v_4 - 2,  v_4,
\end{eqnarray}
where $v_1$, $v_2$ and $v_3$ are three stretching, $v_4$ and $v_5$ are two deformational (asymmetric) bending and $v_6$
is inversion local mode (\TROVE) quantum numbers. The mapping between these quantum numbers for a particular level also required knowledge of the energy value and symmetry, since multiple levels may be labelled with the same local mode quantum numbers. In these ambiguous cases, the symmetric modes quantum numbers $n_1$ and $n_2$ were chosen for the lower energies, and $n_3$ and $n_4$ with the higher, and that $L_3$ and $L_4$ increase proportionally with the energies, and are multiples of 3 in the case of $A_1$ or $A_2$ symmetries, or otherwise for the $E$-type symmetry. This mapping is performed by hand at the $J$ = 0 stage of the calculation, and then propagated to $J >$ 0.

The Einstein-$A$ coefficient for a particular transition from the
\textit{initial} state $i$ to the \textit{final} state $f$ is given by:
\begin{equation}
A_{if} = \frac{8\pi^{4}\tilde{\nu}_{if}^{3}}{3h}(2J_{i} + 1)\sum_{A=X,Y,Z}
|\langle \Psi^{f} | \bar{\mu}_{A} | \Psi^{i} \rangle |^{2},
\end{equation}
where  $h$ is Planck's constant, $\tilde{\nu}_{if}$ is the wavenumber of the line, (\(hc \,
\tilde{\nu}_{if} = E_{f} -E_{i}\)), $J_{i}$ is the rotational quantum number for the initial state, \(\Psi^{f}\) and \(\Psi^{i}\) represent the \TROVE\ eigenfunctions of the final and initial states respectively, \(\bar{\mu}_{A}\)
is the electronically averaged component of the dipole moment along the space-fixed
axis \(A=X,Y,Z\) (see also \citet{05YuThCa.method}).

In order to calculate the absorption intensity for a given temperature $T$
only five quantities are required (all provided by UYT2), the transition wavenumber $\tilde{\nu}_{if}$,
the Einstein coefficient $A_{if}$, the lower (initial) state energy term value $\tilde{E}_{i}$, the total degeneracy
of the upper (final) state $g_{f} = g_{\rm ns} J_f (J_f+1)$, and the partition function $Q(T)$, as given by
\begin{equation}
I(f \leftarrow i) = \frac{A_{if}}{8\pi c}\frac{g_{f}}{Q\;\tilde{\nu}_{if}^{2}}
\exp\left(-\frac{c_2 \tilde{E}_{i}}{T}\right)
\left[1-\exp\left(\frac{-c_2\tilde{\nu}_{if}}{T}
\right)\right] ,
\label{e:intSim}
\end{equation}
where \(c_2\) is the second radiation constant, $g_{\rm ns}$ is the nuclear spin statistical weight factor ($g_{\rm ns}$ = 1 for $^{32}$S$^{16}$O$_{3}$), $c$ is the speed of light.
The partition function $Q$ is given by:
\begin{equation}
  Q = \sum_{i} g_{i}\exp\left({\frac{-c_2 \tilde{E}_{i}}{T}}\right).
  \label{eq:part}
\end{equation}

For a line list to be suitable
for modelling spectra at a certain temperature it is necessary for the
partition function, $Q$, to be converged at this temperature. This is
equivalent
to stating that all energy levels that are significantly populated at the given
temperature, $T$, must be considered. This convergence gives a metric upon which line list
completeness can be gauged \citep{jt181}.

\begin{figure}
\includegraphics[width=\columnwidth]{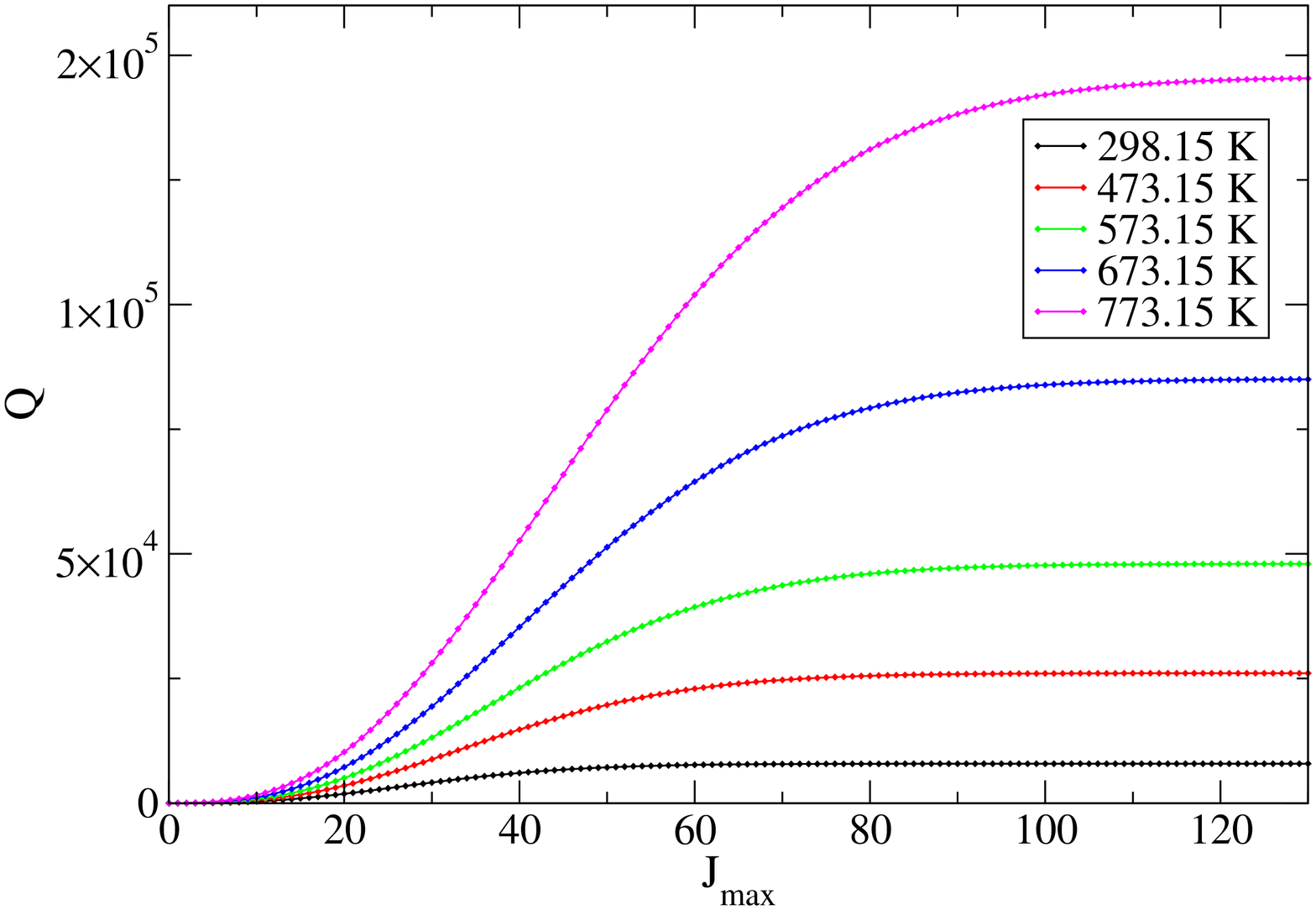}
\includegraphics[width=\columnwidth]{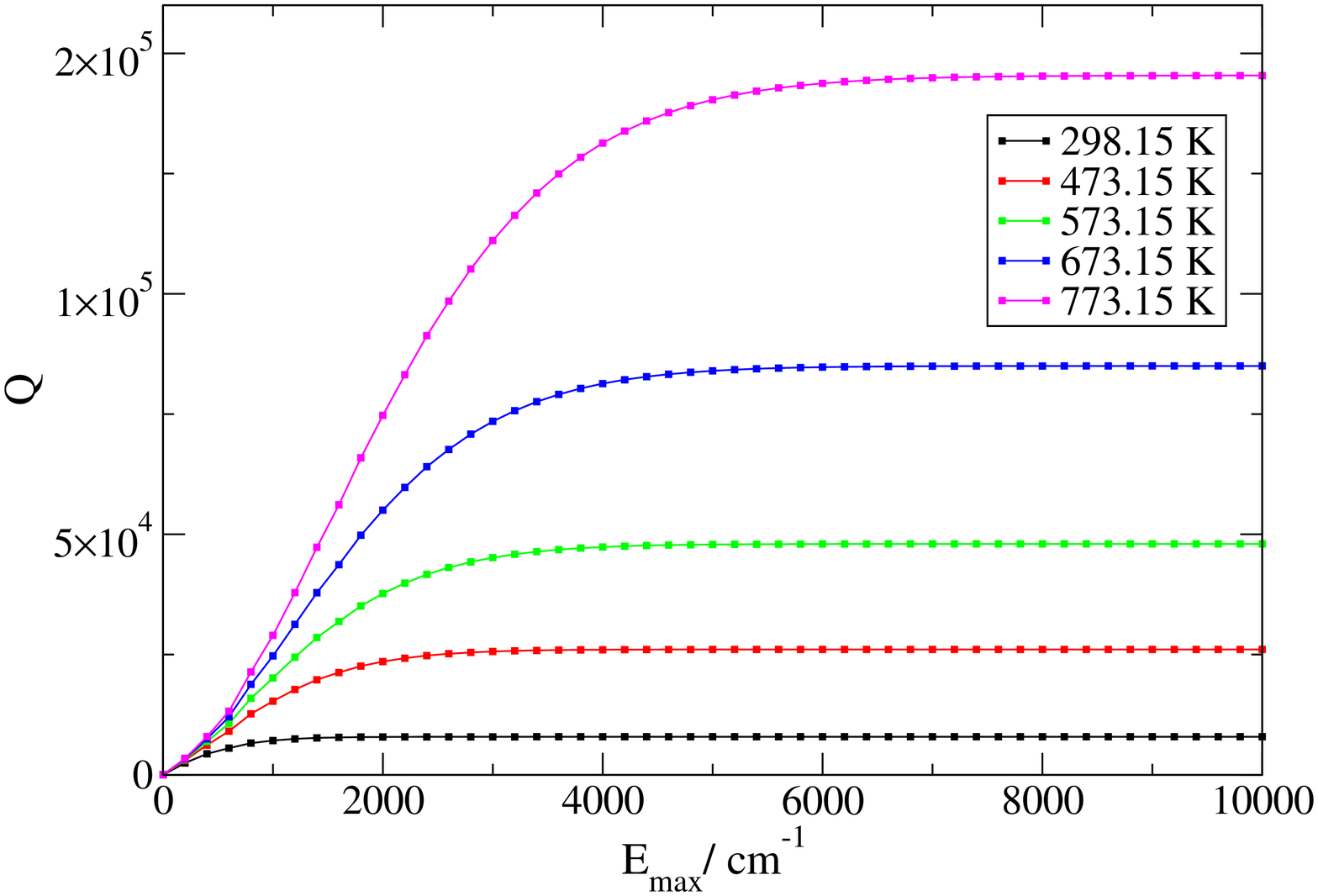}
\caption{Convergence of partition function at different temperatures as a
function of $J_{\rm max}$ (upper) and  $E_{\rm  max}$ (lower). The partition function increases monotonically with temperature. }
\label{so3hotPF}
\end{figure}

Figure \ref{so3hotPF} shows convergence of the partition function with
$J_{\rm max}$ for different temperatures, $T$.
Upon inspection, the value of $Q$ is adequately converged
at $J$ = 130 for $T \leq 800$~K. Table \ref{so3Qconv_temp} shows the final
values of $Q$ obtained for selected temperature  alongside their estimated
degree of convergence. As can be seen, the value of $Q$ = 7908.906 at $T$ =
298.15 K calculated from UYT2 is in agreement with the value of $Q$ =
7908.266 obtained from UYT.

\begin{table}
\caption{Values of the partition function, $Q$, for different temperatures,
$T$. The degree of convergence is specified by $Q_{J_{130}} -
Q_{J_{129}}/Q_{J_{130}} \times$ 100.}
\centering
\begin{tabular}{lrr}
\hline\hline
T (K)& Q & Degree of Convergence (\%)\\
\hline

298.15 & 7908.906 & 6.27 $\times\ 10^{-6}$\\
473.15 & 26065.642 & 8.50 $\times\ 10^{-4}$\\
573.15 & 48007.866 & 3.62 $\times\ 10^{-3}$\\
673.15 & 85016.645 &  9.99 $\times\ 10^{-3}$\\
773.15 & 145389.574 & 2.12 $\times\ 10^{-2}$\\
1000  & 437353.233 \\
\hline
\hline

\end{tabular}
\label{so3Qconv_temp}
\end{table}

For the purposes of determining completeness of the line list, it is
more appropriate to view the convergence of $Q$ as a function of an
energy cut-off, $E_{\rm max}$. This is also shown in Figure
\ref{so3hotPF}, from where it is clear from that imposing this limit
will have a non-negligible effect on a spectral simulation at $T$ =
773.15 K, in particular; since the partition function is not fully
converged at $E_{\rm max}$ = 4000 \cm\ it is expected that levels with
energies above this value will also be populated to some extent. This
would be manifest as certain lines being missing from the spectrum,
where transitions from levels contributing with some significance to
the partition function are not included.  Similarly, the truncation of
calculations at $J$ = 130 means that a number of potentially
contributing energy levels are omitted from the partition sum at $T$ =
773.15 K; at $J$ = 130 the lowest energy lies around 4000 \cm. This
means that the high-$T$ partition function obtained will be slightly
lower than the fully converged value.

It is possible to quantify the
completeness of the line list by assuming that the value of $Q$ at $J$ = 130 is
close enough to the `true' value of the partition function at the given
temperature. Figure \ref{so3hotPFTemp} shows the ratio of the value of the
partition function at the 4000 \cm\ cut-off and the assumed total partition
function, $Q_{\rm Total}$.
At $T$ = 773.15 K, the line list is roughly 90\% complete. In reality, this is
an upper limit due to the fact that there is a slight underestimation of
$Q_{\rm Total}$ at this temperature. However, the contribution from the missing energies
with $J >$ 130, which all lie above
4000 \cm, can be estimated to be small enough
not to affect  $Q_{\rm Total}$ by more than 1\%\ below $T= 800$~K.

\begin{figure}
\includegraphics[width=\columnwidth]{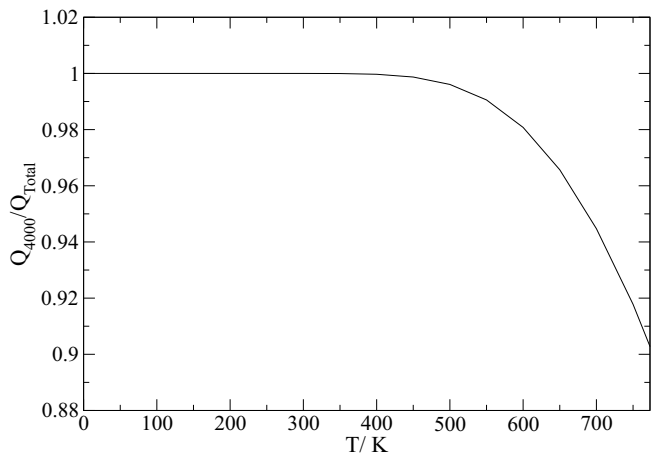}
\caption{Ratios of $Q_{4000}$ to the assumed converged values $Q_{\rm Total}$ as a
function of temperature.}
\label{so3hotPFTemp}
\end{figure}

\section{Intensity Comparisons}

UYT \citep{jt554} made extensive intensity comparisons with the
available, room temperature, high resolution, infrared spectra due to
Maki \etal; in general finding good agreement.  However, these
experimental spectra are not absolute so the comparison is only for
relative intensities.  A comparison of the intensities predicted by
the UYT and UYT2 line lists are summarised in Table
\ref{numLines_hot}.  This comparison essentially shows that UYT2
reproduces the band intensities of UYT, showing that adjusting PES does not
significantly alter the computed intensities, as has occasionally been
found to happen \citep{jt597}.

\begin{table}
\caption{Comparison of calculated band intensities in cm/molecule$\times 10^{-18}$. Units are given in
10$^{-18}$ cm molecule$^{-1}$.}
\centering
\begin{tabular}{lrr}
\hline\hline
Band &  \multicolumn{2}{c}{Band Intensity}\\
& UYT & UYT2\\
\hline
2$\nu_2$ - $\nu_2$ &  0.66 & 0.62\\
$\nu_2$  &  3.71 & 3.39\\
$\nu_2$ + $\nu_4$ -$\nu_4$ & 0.58 &  0.54\\
$\nu_4$  &  5.95 & 5.37\\
2$\nu_4^{(l_4 = 0)}$ - $\nu_4$ & 0.41 &  0.44\\
$\nu_2$ + $\nu_4$ - $\nu_2$ & 0.53 & 0.49\\
2$\nu_4^{(l_4= 2)}$ - $\nu_4$ & 0.87 &  1.17\\
$\nu_1$ - $\nu_4$ & 0.10 &  0.22\\
$\nu_3$  & 44.44 &  43.21\\
2$\nu_3^{(l_3 = 2)} - \nu_0$ & 0.12 & 0.11\\
\hline\hline
\end{tabular}
\label{numLines_hot}
\end{table}

Since the comparison with the data of Maki \etal\ is only able to provide a
measure of the quality of relative intensities within a particular band,
an absolute intensity comparison is highly desirable.
The new measured temperature-dependent DTU
SO$_3$ cross section data plus the
room-temperature cross sections in the PNNL (Pacific Northwest National
Laboratory)
database \citep{sjs04} provide this possibility. For both data sets there
are discernible spectral features across four separate  regions and it should be possible to make a
semiquantitative analysis by comparing integrated intensities across a given
spectral window. To make this comparison cross sections
were generated from UYT2 using the ExoCross tool \citep{jt542,jt631}.

Figure \ref{so3-pnnl-cs-nu2nu4} shows comparisons between recorded cross
sections from PNNL at 298.15 K (25$^{\circ}$C) and resolution 0.112 \cm,
compared with simulated cross
sections using the full UYT2 line list, based on a Gaussian profile of HWHM =
0.1 \cm. Figure \ref{so3-pnnl-cs-2nu3} gives a similar comparison for
the $\nu_1 + \nu_3$  and  2$\nu_3$ bands.

\begin{figure}
\includegraphics[width=\columnwidth]{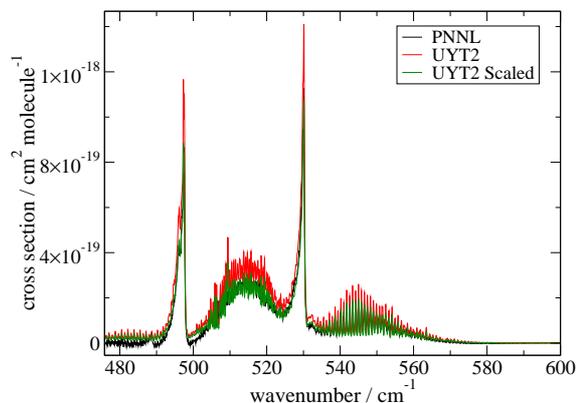}
\includegraphics[width=\columnwidth]{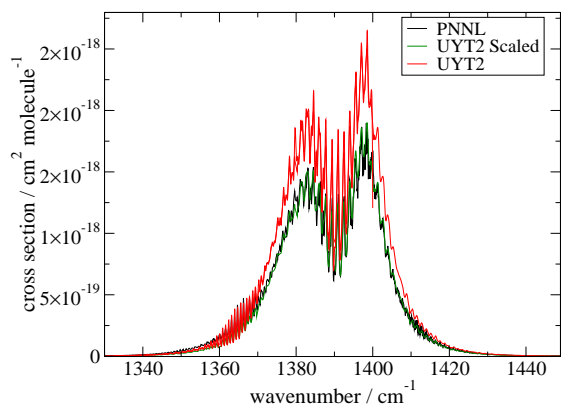}
\caption{Comparisons of the $\nu_2$ and $\nu_4$ bands (upper) and the $\nu_3$
band (lower) for PNNL  \citep{sjs04} and simulated cross sections at $T$ = 298.15 K.}
\label{so3-pnnl-cs-nu2nu4}
\end{figure}
\begin{figure}
\includegraphics[width=\columnwidth]{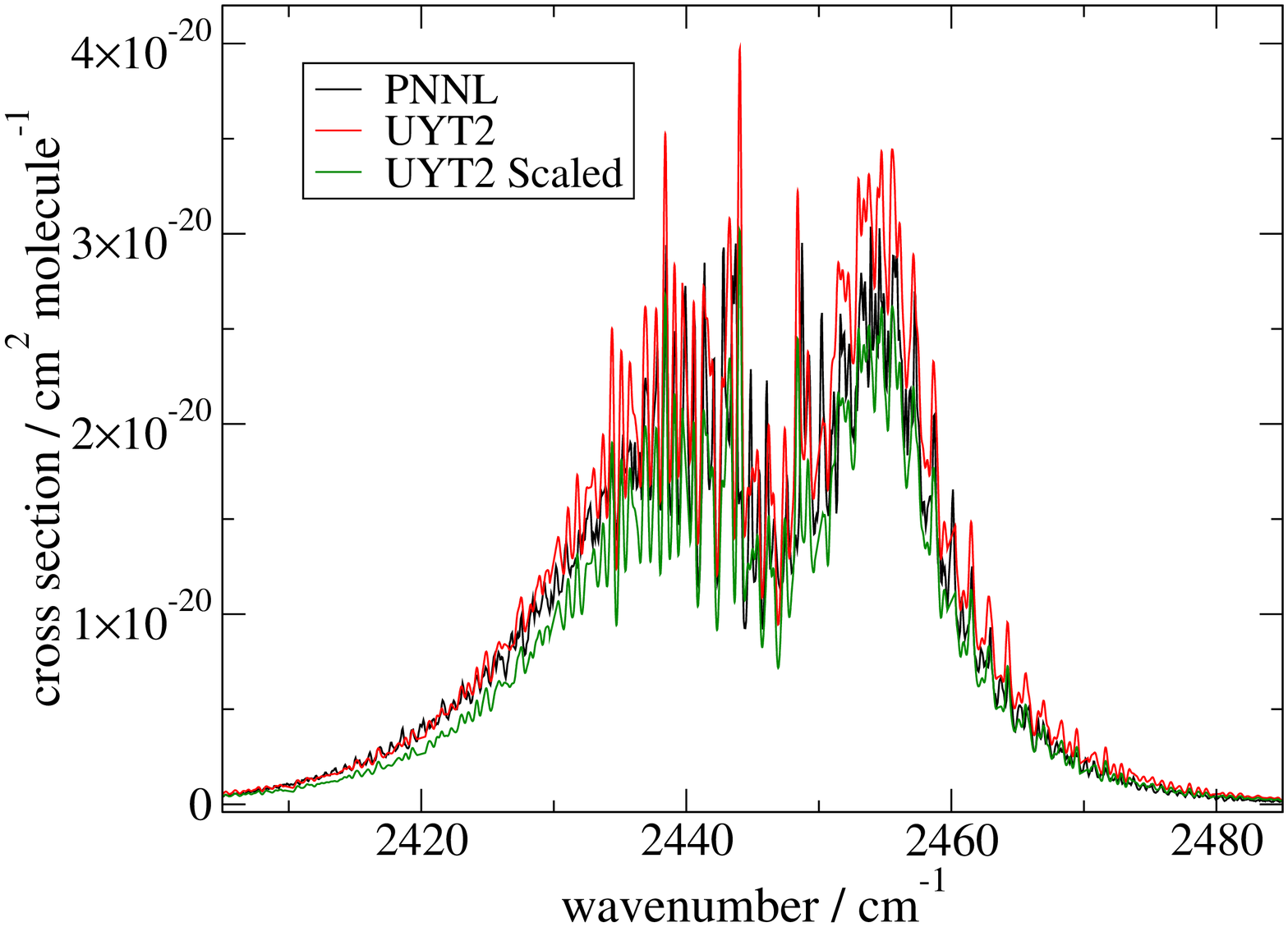}
\includegraphics[width=\columnwidth]{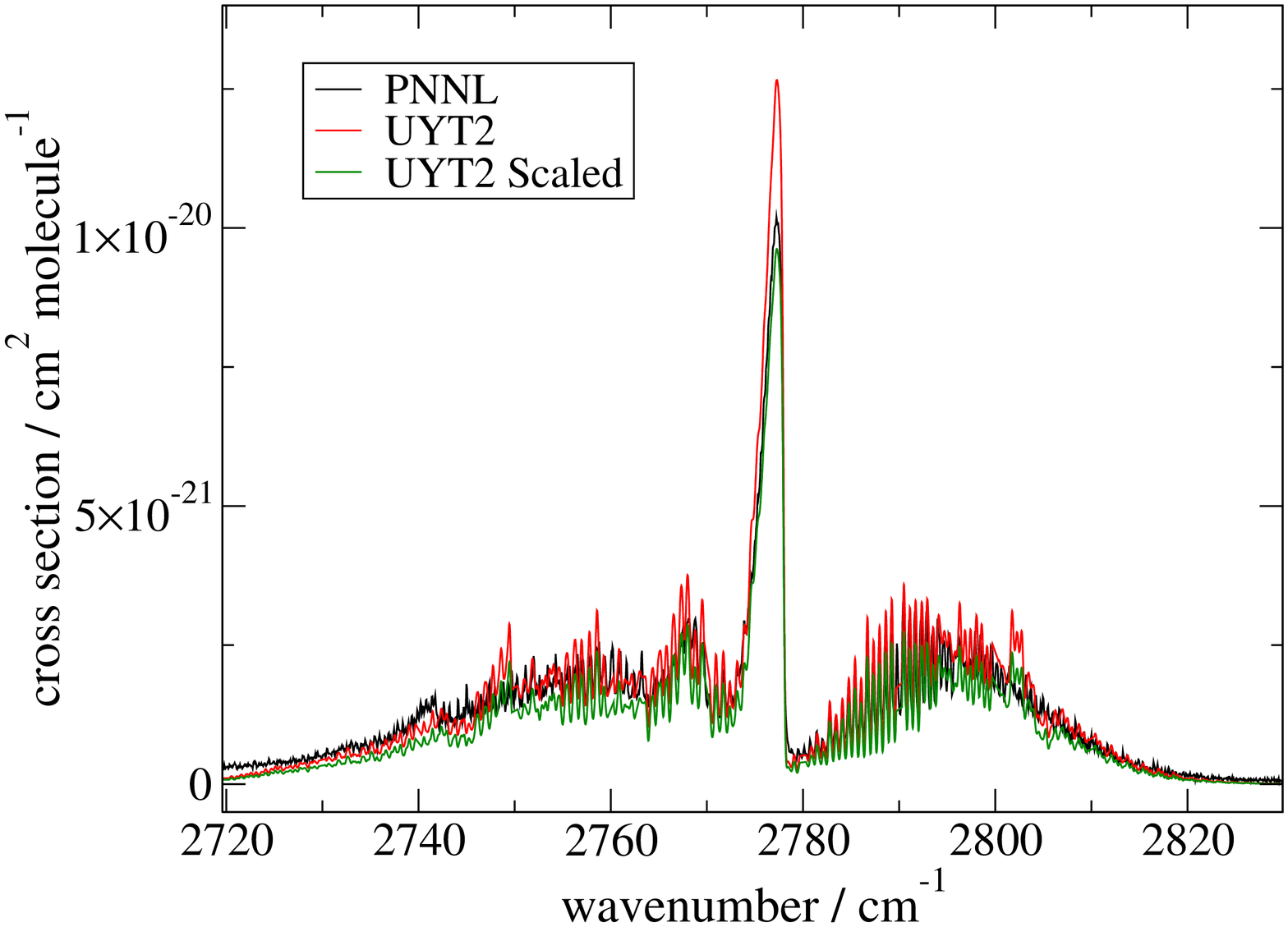}
\caption{Comparisons of the $\nu_1 + \nu_3$ bands (upper) and the 2$\nu_3$
band (lower) for PNNL  \citep{sjs04} and simulated cross sections at $T$ = 298.15 K.}
\label{so3-pnnl-cs-2nu3}
\end{figure}

\begin{figure}
\includegraphics[width=\columnwidth]{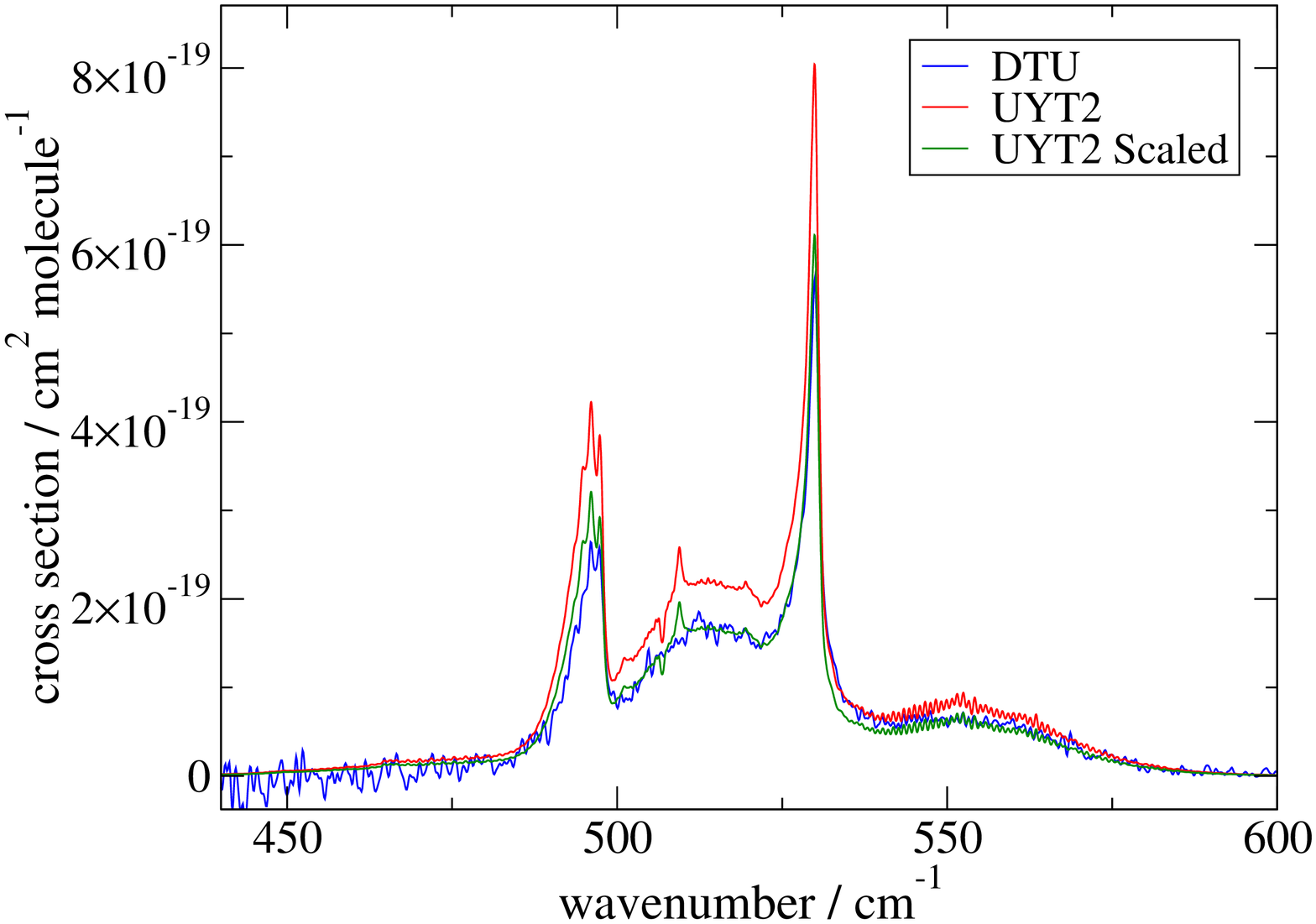}
\includegraphics[width=\columnwidth]{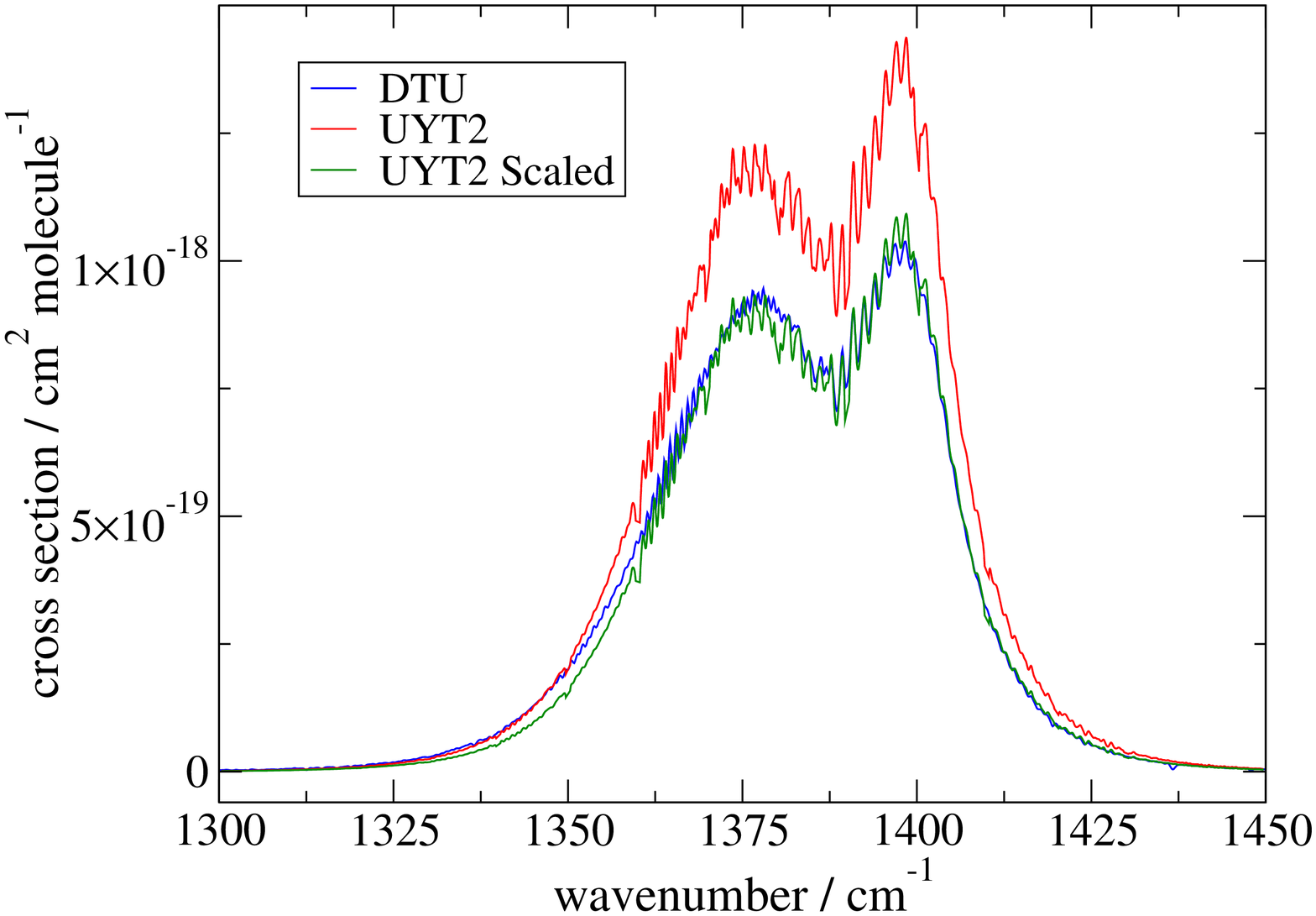}
\caption{Comparisons of the $\nu_2$ and $\nu_4$ bands (upper) and the $\nu_3$
band (lower) for experimentally obtained (this work)  and simulated cross
sections at $T$ = 573.15 K.}
\label{so3-dtu-300-cs-nu2nu4}
\end{figure}

Figure \ref{so3-dtu-300-cs-nu2nu4} shows a comparison of the the $\nu_2$ and
$\nu_4$ complex and the $\nu_3$ band between cross sections recorded for SO$_3$
at 573.15 K (300$^{\circ}$C) and those simulated using the UYT2 line list,
based on a Gaussian profile of HWHM = 0.25 \cm, this  value
is the one which gives  the best representation of the
experimental spectra. In practice, the integrated intensity across the spectral
region is largely independent of the HWHM value used. Figure
\ref{so3-dtu-300-cs-nu1nu3} shows the same comparison for the $\nu_1 + \nu_3$ band, which appears in a noisier region of the spectrum, and is also
disturbed by a strong, foreign absorption feature resulting from the
presence of CO$_2$. There is no data at $T$ = 573.15 K for the
2$\nu_3$ band due to noise contamination in the associated spectral
region. Measurements of SO$_3$ were also made for 773.15 K
(500$^{\circ}$C), however it has not been possible to generate
accurate experimental cross section values due to difficulties in
estimating the concentration within the gas flow cell.  The integrated
absorption cross sections reconstructed from the experimental data at 500$^{\circ}$C
are larger values than those at lower temperatures ($<$
500$^{\circ}$C) suggesting non-conservation of the integrals over the various
SO$_3$ bands. This might be explained by other,
probably hetero-phase processes, which give rise to
different SO$_3$ concentrations than one would expect from the 
assumption made that all SO$_2$ consumed  in the reaction  (\ref{r1}) gives raise of SO$_3$.

\begin{figure}
\includegraphics[width=\columnwidth]{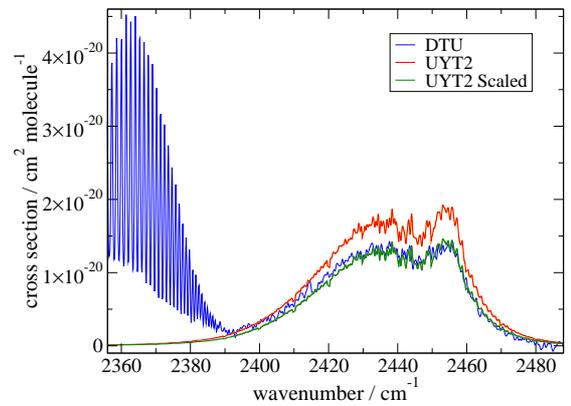}
\caption{Comparisons of the $\nu_1$ + $\nu_3$ bands for experimental
(this work) and simulated cross sections at $T$ = 573.15 K.}
\label{so3-dtu-300-cs-nu1nu3}
\end{figure}

The comparisons reveals that although band positions and features are
fairly well represented, there is  a clear tendency for the UYT2
data to overestimate the
line intensities for both temperatures considered.
In our experience of computing \ai\ intensities it is common for whole bands to
have intensities which are over/underestimated by a constant factor
\citep{jt424}.
However, we have not previously encountered a situation where 
the intensities of all the bands
are shifted by a similar amount.
There are a number of possibilities that
could explain such a discrepancy. Firstly, it is possible that the
experimental cross sections may be underestimated due to an overestimate of the
SO$_3$
abundance; the calculation of cross sections requires the knowledge of the
species concentration within the length of the absorption cell \citep{jt616}.
However,
the fact that measurements at room-temperature performed at DTU corroborate the
PNNL data, and that similar discrepancies are observed for both data sets
suggests that this is not the case. In this context it is worth noting
that a similar comparison for SO$_2$ yields good agreement between
measure and \ai\ absolute cross sections \citep{jt635}.

A second possible source of disagreement could be convergence
issues with the partition function. Since the calculated intensities given by
Eq.~(\ref{e:intSim}) depend on the scaling factor $Q(T)$ the incorrect
computation of this value at the given temperature will lead to inaccurate
values of absolute intensity. The difference in integrated cross section
intensities observed suggest that, if the calculated value of $Q(T)$ is
incorrect, then it is smaller than the `true' value, since the theoretical
cross sections are more intense than the experimentally observed values.
This scenario can also be ruled out, due to two reasons. First, the agreement
between $Q(T)$ for both UYT and UYT2 is very good at $T$ = 298.15 K, where they
are both adequately converged; the increased basis set size underlaying the
UYT2 calculations would undoubtedly account for any missing rovibrational
energies in UYT. Secondly, and perhaps more interestingly, the analysis of
several bands across different temperatures shows the cross section
discrepancies to be almost independent of the value of $T$ (see below). This
would not be expected if $Q(T)$ were the source of the disagreement, since
partition sums can be expected to converge differently as a function of
temperature.

This heavily implies that the problem lies with the DMS defects are by no means unknown
\citep{jt597,jt603,jt607}, despite previous experience of obtaining accurate
\ai\ dipole surfaces \citep{jt509,jt613}.
We therefore undertook a small series of new {\it ab initio} calculations
to see if we could identify the source of this problem. These calculations
were all performed with {\sc MOLPRO} \citep{12WeKnKn.methods}
at the CCSD(T) level using finite differences. First we compared the original
CCSD(T)-F12b with triple-$\zeta$
basis sets (aug-cc-pVTZ-F12 on O and aug-cc-pV(T+d)Z-F12 on S) results
with calculations
performed at the more traditional CCSD(T) with the same basis sets. The
results were very similar suggesting that use of F12b what neither the cause
of the problem nor was it providing improved convergence. Second we
repeated the  CCSD(T) using a larger quadrupole-zeta basis set
(aug-cc-pVQZ-F12 on O and aug-cc-pV(Q+d)Z-F12 on S). The dipoles
computed at this level proved to be somewhat smaller suggesting
that the UYT DMS suffers from a lack of convergence in the one-particle
basis set. Further work on this problem is left to future work.
Here we adopt the more pragmatic approach of scaling our computed
intensities.

It is not easy to make a rigorous analysis based on cross section data
available for SO$_3$, as it is not immediately obvious what the contributions
are from individual lines. In addition to this, both data sets contain varying
degrees of noise within certain spectral regions, with the region around the
$\nu_3$ band generally providing the best signal.
\citet{jt603} performed a fit of their DMS based on experimental intensity data
for nitric acid, to better improve simulated intensities. The lack of absolute
intensity measurements for SO$_3$, coupled with the expensive computational
demands of the line list calculation make this particularly difficult to
perform
here. Nevertheless, the best approach has been to compare integrated band
intensities across fixed spectral windows to obtain scaling parameters for the
each band. Table \ref{so3-cs-comp} summarises the ratios of integrated
intensities between simulated and recorded cross sections for some available
bands.
These were obtained by explicit numerical integration over the 
wavenumber range of the corresponding regions. 

\begin{table}
\caption{Intensities integrated over the corresponding band region for observed and calculated
(UYT2) cross sections as a function of temperature, $T$. Intensity units are given in 10$^{-18}$ cm
molecule$^{-1}$.}
\centering
\begin{tabular}{llrrr}
\hline\hline
$T$ (K) & Band &  \multicolumn{2}{r}{Integrated Intensity}&
Obs./UYT2\\
& & Obs. & UYT2 & \\
\hline
298.15 & $\nu_2\ \&\ \nu_4$ & 9.95 & 13.13 & 0.76\\
 & $\nu_3$ & 46.78 & 60.38 & 0.77\\
 & $\nu_1 + \nu_3$ & 0.71& 0.82 & 0.87\\
 & 2$\nu_3$ & 0.15 & 0.16 & 0.97\\
\hline
573.15 & $\nu_2\ \&\ \nu_4$ & 10.26 & 13.53 & 0.76\\
 & $\nu_3$ & 46.79 & 59.62 & 0.78\\
 & $\nu_1 + \nu_3$ & 0.69 & 0.87 & 0.79\\
\hline\hline
\end{tabular}
\label{so3-cs-comp}
\end{table}

For most bands there appears to be a fairly consistent shift in intensity
values across different temperatures, however the overtone bands for $T$ =
298.15 K suggest otherwise. The differences are quite subtle; for example,
while
the 2$\nu_3$ band has almost perfect agreement in integrated intensity across
the band, the central $Q$-branch peak is not well represented by the UYT2 cross
sections. On the other hand, the DTU data at 573.15 K for the
$\nu_1 + \nu_3$ band exhibits the same general shift as the $\nu_2$, $\nu_3$
and $\nu_4$ bands when care is taken exclude the intensity due to contamination
in the integration, but the same is not true at room-temperature. The PNNL
room-temperature cross sections are presented as a composite spectrum created
from 8 individual absorbance spectra taken at various different pressures using
both a mid-band MCT and wide-band-MCT detector, and uncertainties
in intensity are listed as 10\%. 
The measurements at DTU were performed several times and over different years, when the cell was used for other measurements. The data however are well-reproducible. Up to 400$^{\circ}$C agreement in integrated absorption cross sections between DTU and PNNL is from 0~\% to 13~\% for strongest bands that is the same order as PNNL's uncertainties in bands intensity.
If the scaling factors for the two overtone bands
at room-temperature are ignored, then the remaining factors may be averaged and
applied to all simulated cross sections. This gives an average scaling factor of 0.76.
The assumption made here is that the apparent better agreement in the room-temperature intensities for the $\nu_1
+ \nu_3$ and 2$\nu_3$ are `accidental', while the wide, coverage-consistent
high-temperature cross sections provide a more accurate description of the
differences. Previous experience suggests that an \ai\ DMS is more likely
to overestimate rather than underestimate intensities
\citep{sp00,jt573,jt607}. Without extra experimental data for more bands at
different temperatures it is difficult to ascertain whether the intensity
overestimates seen here are consistent for all bands or vary with
different vibrational transitions.

Figures \ref{so3-pnnl-cs-nu2nu4} and \ref{so3-pnnl-cs-2nu3} show
the various bands at room-temperature, with computed cross sections multiplied
by the averaged scaling factor. Figures \ref{so3-dtu-300-cs-nu2nu4} and
\ref{so3-dtu-300-cs-nu1nu3} show the same for $T$ = 573.15 K, which improve
the simulated cross sections, and demonstrate the implied temperature
independent nature of the discrepancy. As can be seen in Figure
\ref{so3-pnnl-cs-2nu3}, using the averaged scaling factor (obtained from
excluding the individual $\nu_1 + \nu_3$ and 2$\nu_3$ Obs./UYT2 ratios)
improves the reproduction of the central band peak, though the $P$-branch does
show some intensity differences. This appears to be common for multiple bands
and is possibly due to our neglecting of pressure broadening when
generating the cross sections.

Figure \ref{so3-xsec-whole} shows the cross sections calculated over the entire
spectral range of 0 $ < \nu \leq$ 5000 \cm, using a Gaussian profile of HWHM =
0.25 \cm, for a number of different temperatures. All simulated cross sections
have been multiplied by the average scaling factor of 0.76. As can be seen,
the region beyond 4500 \cm\ shows some anomalies for higher temperatures, for
this reason it is
recommended that this region be treated with caution.

\begin{figure}
\includegraphics[width=\columnwidth]{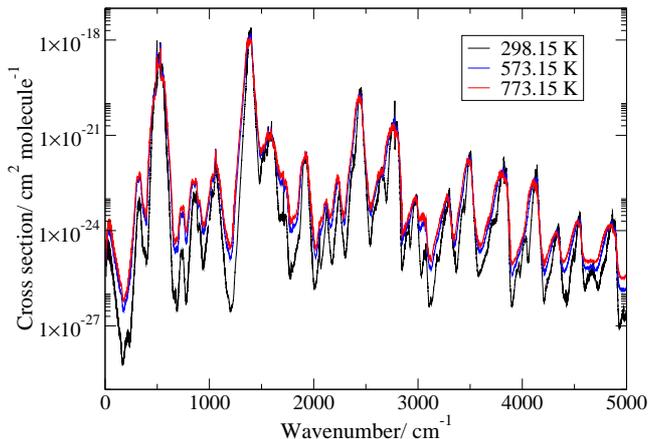}
\caption{Overview of the simulated cross sections using UYT2, at $T$ = 298.15,
473.15 and 773.15 K, with a Gaussian profile of HWHM = 0.25 \cm.
The dips in the cross sections are progressively smoothed out with increasing temperature.
For higher
temperatures, the region beyond 4500 \cm\ appears anomalous, and should be
treated with caution.}
\label{so3-xsec-whole}
\end{figure}

\section{Conclusion}

The UYT2 line list contains 21 billion transitions, and a total of 18 million
energy levels below 10~000\cm. This provides an improvement upon the initial
room-temperature line list, UYT, in terms of both line positions and
temperature coverage.
Table \ref{rmsBands_hot} provides a measure of the improvement introduced by
the PES refinement present in the UYT2 line list. The total RMS deviation for
the bands included in the potential adjustment is 1.35 \cm, compared to 3.23
\cm\ for the unrefined PES of UYT. The majority of simulated line positions
across these bands is improved by an order of magnitude.

It is
difficult to ascertain the overall quality of the \ai\ DMS used in the
production of line intensities. However comparing with newly available cross
section data at two different temperatures heavily suggests that the DMS used
in the calculation of intensities is slightly overestimated, causing an
apparently
constant shift in all intensity values. The evidence suggesting this
temperature- and band-independent scaling factor is certainly not conclusive,
and one may wish to take care in which scaling factor to use for each band. In
particular, bands for which there are no experimental intensity data available
can not be considered to be truly represented well in UYT2 and the lack of
exhaustive absolute intensity knowledge for SO$_3$ limits our ability to
effectively correct for the
disagreements observed. Nevertheless it is hoped that the scaling factor
improves the \ai\ intensity values produced in the UYT2 line list.
Further work,
probably starting with a systematic \ai\ study of the type recently
performed for H$_2$S by \citet{jt607}, is required in order to fully investigate
the source of this
discrepancy.  An experimental determination of individual line intensities
would also be extremely helpful.

The increased size of the basis set, the computation of rovibrational energies
up to $J$ = 130, and the increased spectral range of line strength calculations
allows for UYT2 to be used in the simulation of spectra between 0 $ < \nu \leq$
5000 \cm, with approximately 90\%\ completion at $T$ = 773.15 K (500 C),
however calculated cross sections for the region beyond 4500 \cm\ should be
treated cautiously, and will have to be further investigated. Given that this
is
the largest data set of its kind for \sothree, it is recommended that UYT2 be
used in the production of cross sections at room-temperature, and up to $T$ =
773.15 K, for both astronomical and other applications.

The UYT2 line list contains 21 billion transitions. This makes it use
in radiative transport modelling computationally challenging. Work on
an even larger methane line list \citep{jtCH4} suggests that it should
be possible to split the list into a temperature-dependent but
pressure-independent, background cross section which is used to augment a hugely
reduced list of stong line whose profiles are treated in detail.
This idea will be explored further in the future.

\section*{Acknowledgements}

This work was supported by Energinet.dk project 2010-1-10442 \lq\lq
Sulfur trioxide measurement technique for energy systems'' and the ERC
under the Advanced Investigator Project 267219. It made use of the
DiRAC@Darwin HPC cluster which is part of the DiRAC UK HPC facility
for particle physics, astrophysics and cosmology and is supported by
STFC and BIS as well as the Emerald Computing facility funded by
EPSRC.
We thank Cheng Liao and the SGI team for providing the Plasma diagonaliser and DiRAC project at
Cambridge for assistance in running our jobs.

\bibliographystyle{mnras}
\bibliography{journals_astro,jtj,SO2,SO3,linelists,methods,additional,abinitio,exogen,NH3,CO2,phdthesis,h2s,SO,H2CO,DMS,exoplanets,H2O,PAH,diatomic}
\label{lastpage}

\end{document}